\documentclass[pre,aps,floatfix,notitlepage]{revtex4-1}

\usepackage[dvips]{graphicx}
\usepackage{amssymb,amsfonts,amsmath,calc,ifsym,bbding,wasysym}
\usepackage[usenames,dvipsnames]{xcolor}
\usepackage{bm}
\usepackage[normalem]{ulem}
\usepackage{xr}
\usepackage{float}

\def\eunit{E_\text{u}}

\def\dunit{D_\text{u}}
\def\tunit{t_\text{u}}

\def\qp{q_\text{p}}

\def\rcut{r_\text{cut}}
\newcommand{\sub}[2]{ _{\mathrm{#1}#2}}
\newcommand{\rsub}[1]{_\mathrm{#1}}
\newcommand{\kt}{k_\mathrm{B}T}
\newcommand{\kd}{K_\mathrm{d}}
\newcommand{\sbb}{s_\mathrm{b}}
\newcommand{\scc}{s_\mathrm{c}}
\newcommand{\gcc}{g_\mathrm{cc}}
\newcommand{\uchain}{\mu_\mathrm{chain}}
\newcommand{\ur}{\mu_\mathrm{r}}
\newcommand{\np}{N_\mathrm{p}}
\newcommand{\ui}{U_\mathrm{I}}
\newcommand{\LJ}[1]{ \mathcal{L}_{#1} }
\newcommand{\Morse}[1]{ \mathcal{M}_{#1} }

\newcommand{\Coulomb}[1]{ \mathcal{C}_{#1} }
\def\Lopt{L^{*}_\text{eq}}
\def\Loptk{L^{*}_\text{dyn}}
\def\Csalt{C_\text{salt}}
\def\ld{\lambda_\text{D}}
\def\lp{l_\text{p}}
\def\Rg{R_\text{G}}
\def\nnuc{n_\text{nuc}}

\def\inrad{R_\text{in}}
\def\lpair{l_\text{pair}^\text{negative}}
\def\Lc{L_\text{C}}
\def\Lbp{L_\text{bp}}
\def\fbp{f_\text{bp}}
\def\lb{\lambda_\text{B}}

\begin{document}

\title{Viral genome structures are optimal for capsid assembly}
\author{Jason D Perlmutter}
\author{Cong Qiao}
\author{Michael F Hagan}
\affiliation{Martin Fisher School of Physics, Brandeis University,
  Waltham, MA, USA.}
\maketitle

\section{abstract}
Understanding how virus capsids assemble around their nucleic acid (NA) genomes could promote efforts to block viral propagation or to reengineer capsids for gene therapy applications. We develop a coarse-grained model of capsid proteins and NAs with which we investigate assembly dynamics and thermodynamics. In contrast to recent theoretical models, we find that capsids spontaneously `overcharge'; i.e., the negative charge of the NA exceeds the positive charge on capsid. When applied to specific viruses, the optimal NA lengths closely correspond to the natural genome lengths. Calculations based on linear polyelectrolytes rather than base-paired NAs underpredict the optimal length, demonstrating the importance of NA structure to capsid assembly. These results suggest that electrostatics, excluded volume, and NA tertiary structure are sufficient to predict assembly thermodynamics and that the ability of viruses to selectively encapsidate their genomic NAs can be explained, at least in part, on a thermodynamic basis.

\section{Introduction}
For many viruses the spontaneous assembly of a protein shell, or capsid, around the viral nucleic acid (NA) is an essential step in the viral lifecycle. Identifying the factors which enable capsids to efficiently and selectively assemble around the viral genome could identify targets for new antiviral drugs that block or derail the formation of infectious virions. Conversely, understanding how assembly depends on the NA and protein structure would guide efforts to reengineer capsid proteins and human NAs for gene therapy applications. From a fundamental perspective, high-order complexes that assemble from protein and/or NAs abound in biology. Learning how the properties of viral components determine their co-assembly can shed light on assembly mechanisms of a broad array of structures and the associated selective pressures on their components. In this article, we use GPU computing \cite{Anderson2008,nguyen2011,LeBard2012} and a simplified, but quantitatively testable, model to elucidate the effects of electrostatics, capsid geometry, and NA tertiary structure on assembly.

Assembly around NAs is predominately driven by electrostatic interactions between NA phosphate groups and basic amino acids, often located in flexible tails known as arginine rich motifs (ARMs) (e.g., \cite{ Schneemann2006}). There is a correlation between the net charge of these protein motifs and the genome length for many ssRNA viruses \cite{Hu2008a, Belyi2006}, with a `charge ratio' of negative charge on NAs to positive charge on proteins typically of order 2:1 (i.e., viruses are `overcharged'). Electrophoresis measurements confirm that viral particles are negatively charged (e.g. \cite{Serwer1995,Serwer1999,Porterfield2010}), though these measurements include contributions from the capsid exteriors \cite{LosdorferBozic2012, Zlotnick2013}. Based on these observations, it has been proposed that viral genome lengths are thermodynamically optimal for assembly, meaning that their lengths minimize the free energy of the assembled nucleocapsids. However, while estimates of optimal lengths have varied \cite{Hu2008a,Belyi2006, Schoot2005, Angelescu2006, siber2008, Ting2011, Ni2012}, recent theoretical models based on linear polyelectrolytes \cite{siber2008, Ting2011, Ni2012} have consistently predicted that optimal NA lengths correspond to `undercharging' (fewer NA charges than positive capsid charges). These results lead to the conclusion that capsid assembly around genomic (overcharged) NAs requires an external driving force such as a Donnan potential \cite{Ting2011}. Yet, viruses preferentially assemble around genomic length RNAs even \emph{in vitro} \cite{Comas-Garcia2012}, in the absence of such a driving force.

The effect of NA structural features other than charge remains unclear. In some cases, genomic NAs are preferentially packaged over others with equivalent charge \cite{Borodavka2012} due to virus-specific packaging sequences \cite{Bunka2011, Borodavka2012}. However, experiments on other viruses have demonstrated a striking lack of virus-specific interactions \cite{Comas-Garcia2012, Porterfield2010}. For example, cowpea chlorotic mottle virus (CCMV) proteins preferentially encapsidate BMV RNA over the genomic CCMV RNA \cite{Comas-Garcia2012}. Since the two NAs are of similar length, the authors propose that other structural features, such as NA tertiary structure \cite{Yoffe2008}, may drive this preferential encapsidation. However, the relationship between NA structure and assembly has not been explored.

To clarify this relationship, we use a computational model to investigate capsid assembly dynamics and thermodynamics as functions of NA and capsid charge, solution ionic strength, capsid geometry, and NA size (resulting from tertiary structure). We first test the proposed link between the thermodynamic optimum length, $\Lopt$ and assembly, finding that the yield of assembled nucleocapsids at relevant timescales is maximal near $\Lopt$. Longer-than-optimal NAs lead to non-functional structures, indicating that the thermodynamic optimum ($\Lopt$) corresponds to an upper bound for the genome size for capsids which spontaneously assemble and package their genome. We then explore how $\Lopt$ depends on solution conditions and the structures of capsids and NAs. We find that overcharging occurs spontaneously, requiring no external driving force.  When base-pairing is accounted for, predicted optimal NA lengths are consistent with the genome size for a number of viruses, suggesting that electrostatics and NA tertiary structure are important factors in the formation and stability of viral particles. Our predictions can be tested quantitatively in \emph{in vitro} packaging experiments (e.g. \cite{Comas-Garcia2012, Cadena-Nava2012, Porterfield2010}).

\section{Model}
\begin{figure}
\centering{\includegraphics[width=0.5\columnwidth]{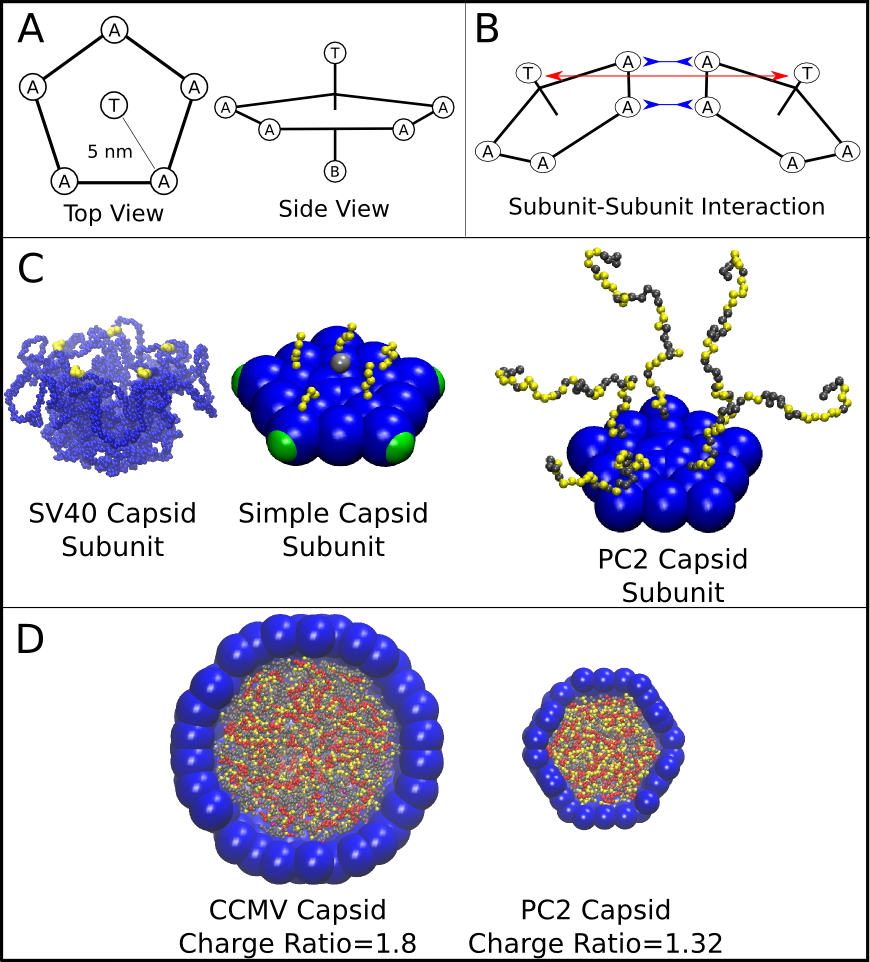}}
\caption{{\bf (A),(B)} Model schematic for (A) a single subunit, and (B) two interacting subunits, showing positions of the attractor (`A'), Top (`T'), and Bottom (`B') pseudoatoms, which are defined in the Model section and in SI section~\ref{sec:S1A}. {\bf (C)} (left) The pentameric SV40 capsid protein subunit, which motivates our model. The globular portions of proteins are shown in blue and the beginning of the NA binding motifs (ARMs) in yellow, though much of the ARMs are not resolved in the crystal structure \cite{Stehle1996}. Space-filling model of the basic subunit model (middle) and a pentamer from the PC2 model (right). {\bf (D)} A cutaway view of complete CCMV and PC2 capsids (with respective biological charge ratios of 1.8 and 1.32). Beads are colored as follows: blue=excluders, green=attractors, yellow=positive ARM bead, gray=neutral ARM bead, red=polyelectrolyte.
\label{fig1}
}
\end{figure}

Our coarse-grained capsid model (Figure~\ref{fig1}) is motivated by the recent observation \cite{Kler2012} that purified simian virus 40 (SV40) capsid proteins assemble around ssRNA molecules \textit{in vitro} to form capsids comprising 12 homopentamer subunits. We model the capsid as a dodecahedron, composed of 12 pentagonal subunits (each of which represents a rapidly forming and stable pentameric intermediate, which then more slowly assembles into the complete capsid, as is the case for SV40 \cite{li2002}). Our model extends those of Refs. \cite{Wales2005, Fejer2009, Johnston2010}, with subunits attracted to each other via attractive pseudoatoms at the vertices (type `A') and driven toward a preferred subunit-subunit angle by repulsive `Top' pseudoatoms (type `T') and `Bottom' pseudoatoms (type `B') (see Fig.~\ref{fig1}, Methods, and SI section~\ref{sec:S1A}). In contrast to previous models for polyelectrolyte encapsidation \cite{Angelescu2006, Elrad2010, Mahalik2012, Kivenson2010}, the proteins contain positive charges located in flexible polymeric tails, representing the ARM (arginine-rich motif) NA binding domains typical of positive-sense ssRNA virus capsid proteins.

To investigate the effect of NA properties on assembly we consider two models for the packaged polymer: (1) a linear flexible polyelectrolyte and (2) a NA with predefined secondary and tertiary structure (i.e. static base-pairs) that captures the size, shape, and rigidity of NAs. Single-stranded polymers are modeled as flexible polymers with one bead per nucleotide \cite{Zhang2004a,ElSawy2011}, with charge $-e$. Double-stranded regions of NAs comprise two adjoined semiflexible strands with the net persistence length of dsDNA ($\approx50$ nm), and base-paired nucleotides are connected by harmonic bonds. Electrostatics are modeled using Debye-Huckel interactions to account for screening, except where these are tested against simulations with Coulomb interactions and explicit salt ions (Fig.~\ref{fig3}D and SI Fig.~\ref{figS1}).

In addition to representing the secondary structures of specific ssRNA genomes, we are able to tune statistical measures of base-pairing, such as the fraction of nucleotides that are base-paired, the relative frequency of hairpins and higher-order junctions (Fig.~\ref{figS3}), and the maximum ladder distance (MLD), which measures the extension in graph space of a NA secondary structure \cite{Yoffe2008}. As shown in Fig.~\ref{figS3}, the radius of gyration $\Rg$ of the model NAs depends on MLD as $1.7\times\text{MLD}^{0.43}$, which has a slightly smaller exponent than a theory in which only base-paired segments were accounted for \cite{Yoffe2008}. Further model details and parameters are presented in the SI.

\section{Results}

\subsection{Capsid assembly leads to spontaneous overcharging}

We begin by presenting the results of simulations on our simplest capsid and cargo models. Our model capsid has a dodecahedron inradius (defined as the distance from the capsid center to a face center) of $\inrad=7.3$ nm, to give an interior volume consistent with that of the smallest icosahedral viruses, and contains 60 ARMs (i.e. a $T{=}1$ capsid, where $T$ is the triangulation number \cite{Caspar1962}) each containing 5 positively charged residues. The cargo is a linear polyelectrolyte. While we systematically alter both the cargo and capsid below to include more biological detail, the simple model demonstrates two important results (that are consistent with results from more complex models): 1) Viral particles spontaneously overcharge during assembly, and 2) The thermodynamic optimal polyelectrolyte length closely correlates with the length for which dynamical assembly leads to the highest yield of complete viral particles.

\begin{figure}
\centering{\includegraphics[width=0.5\columnwidth]{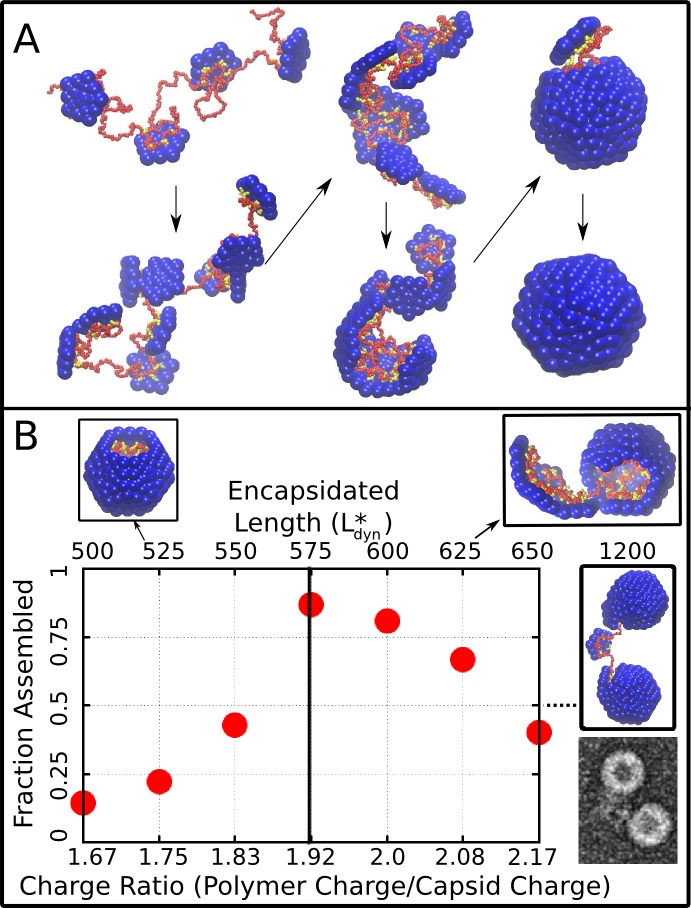}}
\caption{{\bf (A)} Snapshots illustrating assembly of subunits with ARM length=5 around a linear polyelectrolyte with 600 segments. Beads are colored as in Fig.~\ref{fig1}. {\bf (B)} Fraction of trajectories leading to a complete capsid as a function of polymer length (top axis) or charge ratio (bottom axis). The dashed line indicates the thermodynamic optimum charge ratio or length ($\Lopt$) from equilibrium calculations. Snapshots of typical outcomes above and below the optimal length are shown. (Far right) A typical assembly outcome for polymer length 1200 (twice $\Lopt$) is compared to an EM image of CCMV proteins assembled around an RNA which is twice the CCMV genome length \cite{Cadena-Nava2012} (image provided by C. Knobler, W. Gelbart, and R. Garmann).
\label{fig2}
}
\end{figure}

\emph{Dynamical simulations.} The results of Brownian dynamics simulations of capsid assembly around a linear polyelectrolyte at physiological salt concentration (Debye screening length $\ld=1$ nm) are shown in Fig.~\ref{fig2}. Consistent with most ssRNA virus proteins, the polymer is essential for assembly under the simulated conditions, since the subunit-subunit interactions are too weak for formation of empty capsids (see SI section~\ref{sec:FreeEnergy} for details). Fig.~\ref{fig2}A presents representative snapshots of the assembly process for a polyelectrolyte with 600 segments (see also supplemental movies \cite{webnote}). The subunits first adsorb onto the polymer in a disordered fashion, with on average about eight subunits adsorbing before first formation of a critical nucleus (a complex comprising 5 subunits, SI section~\ref{sec:kinetics}). Once a critical nucleus forms, additional subunits add to it sequentially and reversibly until the final subunit closes around the polymer.

The assembly outcome depends on polymer length, with successful capsid formation occurring when there is overcharging, meaning that the negative charge on the polymer exceeds the net positive charge on an assembled capsid ($300e$ for this model). Fig.~\ref{fig2}B shows the yield of well-formed capsids at $t{=}2\times10^{4} \tunit$ ($2\times10^{8}$ time steps), at which point the fraction assembled has approximately plateaued for most parameter values. Here a well-formed capsid is defined as a structure comprising 12 subunits that each interact with five neighboring subunits and together completely encapsulate the polymer. Assembly is robust (yield $\sim0.9$) near an optimal polymer length of $\Loptk {=} 575$ segments, corresponding to a `charge ratio' of $575/300=1.9$. Above the optimal length, the polymer is typically not fully incorporated when capsid assembly nears completion. For sufficiently long polymers (e.g. 2 $\Loptk$, Fig.~\ref{fig2}B right) multiple capsids assemble on the same polymer. These multiplet structures resemble configurations seen in a previous simulation study which did not explicitly consider electrostatics \cite{Elrad2010} and observed in experiments in which CCMV proteins assembled around RNAs longer than the CCMV genome length \cite{Cadena-Nava2012}. For polymer lengths well below $\Loptk$ the polymer is completely encapsulated before assembly completes, and addition of the remaining subunits slows substantially. Although capsids which are incomplete at the conclusion of these simulation might eventually reach completion, the low yield of assembled capsids at our finite measurement time reflects the fact that assembly at these parameters is less efficient than for polymer lengths near $\Loptk$.

\emph{Equilibrium calculations.} We calculated the thermodynamic optimal polymer length $\Lopt$, or the length of encapsulated polymer that minimizes the free energy of the polymer-capsid complex, with two different methods. First, we performed Brownian dynamics simulations of a long polymer and a preassembled capsid with one subunit made permeable to the polymer, so that the length of encapsulated polymer is free to equilibrate. Second, we calculated the residual chemical potential difference between encapsidated polymer and a polymer free in solution \cite{Kumar1991, Elrad2010}. The first method predicts an optimal polymer length of $\Lopt{=}574$ while the latter suggests $\Lopt\approx550-575$, indistinguishable from the optimal length found in the finite-time dynamical assembly simulations (Fig.~\ref{fig2}B). The observation that the yield of encapsulated polymers from dynamical assembly trajectories diminishes above $\Lopt$, together with the observation that many viruses with single-stranded genomes assemble and package their nucleic acid spontaneously, suggests that this equilibrium value may set an upper bound on the size of a viral genome.

\subsection{The effect of control parameters on packaged lengths}

\begin{figure}
\centering{\includegraphics[width=0.9\columnwidth]{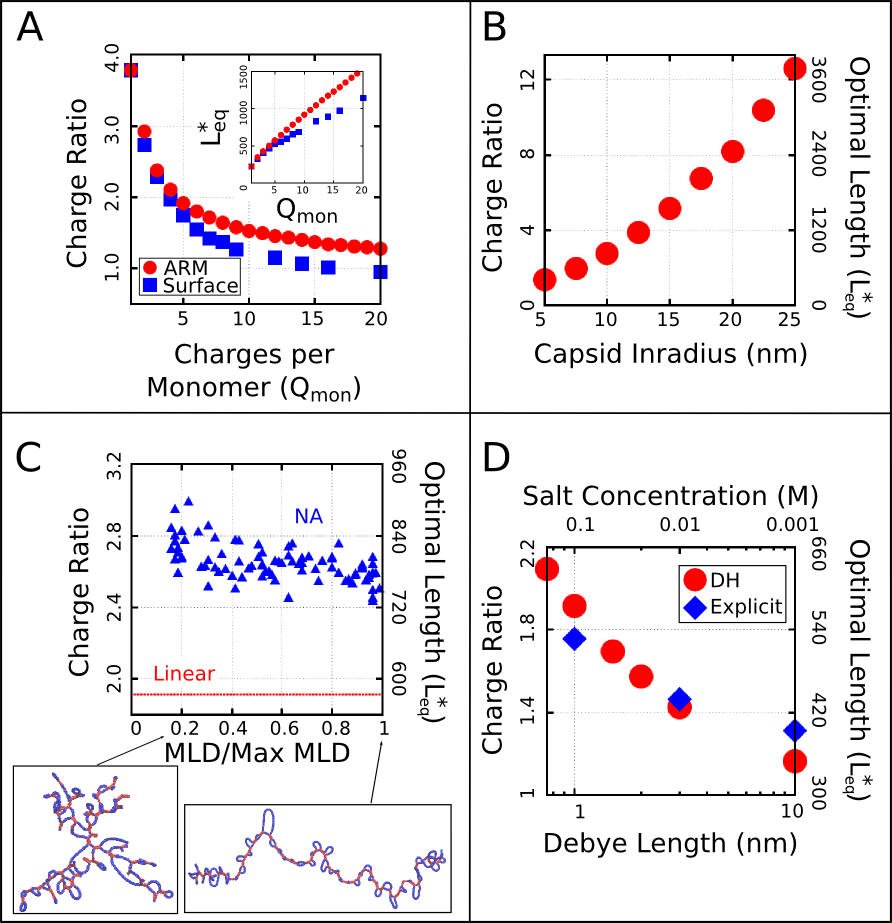}}
\caption{Effect of control parameters on the thermodynamic optimal length and charge ratio. {\bf (A)} Effect of increasing capsid charge, with capsid $\inrad=7.3$ nm. {\bf (B)} Effect of increasing capsid size for fixed ARM length=5. {\bf (C)} Effect of base-pairing, with $\fbp=0.5$ base-paired nucleotides and varying maximum ladder distance (MLD), for capsid $\inrad=7.3$ nm and ARM Length=5. Snapshots of our model NA structures with small and large MLD's are shown (prior to encapsidation), with double-stranded regions in red and single-stranded regions in blue. The result for no base-pairing (linear) is shown as a dashed line.  {\bf (D)} Effect of ionic strength and comparison between Debye-Huckel interactions and explicit ions. The thermodynamic optimum lengths $\Lopt$ and corresponding optimal charge ratios are shown as functions of the ionic strength (Debye screening length), calculated with simulations using Debye-Huckel (DH) interactions (\textcolor{red}{{\Large$\bullet$}} symbols) or Coulomb interactions with explicit ions (\textcolor{blue}{\DiamondSolid} symbols).
\label{fig3}
}
\end{figure}

Since our simulations show that $\Lopt$ and $\Loptk$ are closely correlated, we performed a series of equilibrium calculations in which ionic strength (SI Fig.~\ref{figS1}), capsid structural parameters, and the NA model were systematically varied, to determine the effect of each parameter on $\Lopt$. All results shown in the main text were obtained for physiological ionic strength (1 nm Debye length). To determine how $\Lopt$ and the optimal charge ratio depend on the number of positive charges in the capsid, we first varied the length of the ARMs, keeping all ARM residues positively charged. As shown in Figure~\ref{fig3}A (inset), $\Lopt$ increases sub-linearly with capsid charge, meaning that each additional ARM charge increases the equilibrium polymer packaging length by a smaller amount, leading to a diminishing charge ratio. We obtained a similar result when, instead of modeling flexible ARMs, we placed charges in rigid patches on the inner capsid surface (e.g., corresponding to MS2 \cite{Valegard1997}). However, we find that charges on the surface lead to a lower optimal charge ratio than the equivalent number of charges located in flexible ARMs (Figure~\ref{fig3}A), since the ARM flexibility increases the volume of configuration space available for NA-ARM interactions.

These observations demonstrate that, while electrostatics is an important factor, excluded volume and the lengths of polyelectrolyte segments that bridge between ARMs (discussed below) also affect the length of packaged polyelectrolyte. However, in the biologically relevant range of 5-20 positive charges per protein monomer \cite{Hu2008a, Belyi2006}, the optimal length appears roughly linear with capsid charge (but with a positive intercept).

To understand how capsid size influences $\Lopt$, we varied the model capsid radius while holding the number of capsid charges fixed. As shown in Figure~\ref{fig3}B, $\Lopt$ and hence the optimal charge ratio increase dramatically with capsid size, scaling roughly with capsid radius as $\Lopt \sim \inrad^{1.6}$. The non-integer exponent is intriguing, as it rules out scaling with capsid volume, surface area, or a linear path length, which would respectively result in $\Lopt\sim \inrad^3$, $\inrad^2$, or $\inrad$. Projecting the density of packaged polymer segments onto angular coordinates (Fig.~\ref{figS6}) reveals that the polymer is not homogenously distributed throughout the capsid surface, but instead has enriched density at the vertices and edges relative to the subunit faces. This result is consistent with experimental observations that nucleic acids form dodecahedral cages in viral particles \cite{Speir2012}, and our model may describe scaling of the optimal charge ratio with volume for these capsids. For model capsids with $\inrad \ge 12.5$ nm, the amount of polymer segments directly interacting with ARM charges becomes independent of capsid size, and the dependence of optimal length on volume can be attributed to the lengths of polymer between ARMs (see Discussion and SI section~\ref{sec:bridging}).

{\bf Base-pairing increases packaged lengths.} To understand how the geometric effects of base-pairing contribute to packaging, we performed dynamical assembly simulations and equilibrium calculations of $\Lopt$ for a wide range of base-pairing patterns and fraction of base-paired nucleotides (see SI section~\ref{sec:S1B}). The key result is that for all simulated base-pairing patterns, increasing the fraction of base-paired nucleotides (up to the biological fraction of 50\%) increases $\Lopt$ (Figs.~\ref{fig3}C and ~\ref{figS3}). The increase in optimal length can be as large as 200-250 nucleotides for our small $T{=}1$ capsid, indicating that base-pairing can contribute significantly to the amount of polymer that can be packaged. This effect can be explained by the fact that nucleotide-nucleotide interactions which drive NA structure formation effectively cancel some NA charge-charge repulsions and result in NA structures that are compact in comparison to linear polymers with the same lengths. Thus encapsulated NAs incur smaller excluded volume interactions, electrostatic repulsions, and conformational entropy penalties during assembly.

However, the connection between the size of a molecule in solution and $\Lopt$ is surprisingly subtle. As described in SI section~\ref{sec:S1B}, we have quantified base-pairing patterns by their maximum ladder distance (MLD), which counts the maximum number of base-pairs along any non-repeating path across the NA and thus describes the extent of the molecule in the secondary structure graph space. As shown in Figure~\ref{figS3}, for a NA with 1,000 segments and 50\% base-pairing, the solution radius of gyration varies with MLD as $\Rg \sim \text{MLD}^{0.43}$ to yield $\Rg\approx8$ nm to $\Rg \approx20$ nm, in comparison the linear model $\Rg=25.5$ nm. As shown in Figure~\ref{fig3}C the inclusion of base-pairing has a large effect on $\Lopt$, but changes in MLD have only a minor effect. Though over this range of MLDs the solution $\Rg$ more than doubles, $\Lopt$ changes by only about 10\%, with an even smaller variation over the range of MLDs that we estimate for biological RNA molecules $\text{MLD/Max MLD}\in (0.25,0.55)$ based on Ref. \cite{Gopal2012} (see SI section~\ref{sec:S1B} for additional detail).

{\bf Effect of salt concentration on encapsidation.} As shown in Fig.~\ref{fig1}D, the optimal length $\Lopt$ and charge ratio increase with increasing salt concentration $\Csalt$ (i.e. decreasing Debye length $\ld$) for $Csalt \lesssim 400$ mM. Importantly, the simulations predict overcharging at all salt concentrations investigated ($1\mbox{ mM} \le \Csalt \le 400\mbox{ mM}$). Results from simulations with explicit monovalent ions and those with implicit screening (Debye-Huckel interactions) are also compared in the figure; we see that the two methods of calculating charge interactions agree to within 10\% at the physiological salt concentration (100 mM). We focus on $\Csalt=100$ mM for all other results in this article. Finally, as discussed further in SI section~\ref{sec:salt}, incorporating divalent cations into our model increases $\Lopt$ (as compared to calculations with only monovalent explicit ions), but does not change significantly for experimentally relevant ion concentrations.

\subsection{Predictions for specific viral capsid structures}
To evaluate the significance of the trends identified above for packaging in a biological context, we performed equilibrium calculations in which the structural parameters discussed above (capsid volume, ARM length, charge, and NA base-pairing) were based on specific $T{=}1$ and $T{=}3$ viruses (whose capsids are assembled from 60 and 180 protein copies respectively). For each investigated virus, the capsid radius was fit to protein densities in capsid crystal structures \cite{Carrillo-Tripp2009}, the ARM length was determined from the structure, and charges in the ARM and on the capsid inner surface were assigned based on amino acid sequence (see Table~\ref{tab1}). NAs were modeled with 50\% base-pairing and $\text{MLD/Max MLD}\approx 0.5$. Visualizations of $T{=}1$ and $T{=}3$ viruses (PC2 and CCMV) are presented in Figure~\ref{fig1}D and further details details are provided in SI section~\ref{sec:specificCapsids}.

\begin{figure}
\centerline{\includegraphics[width=0.99\columnwidth]{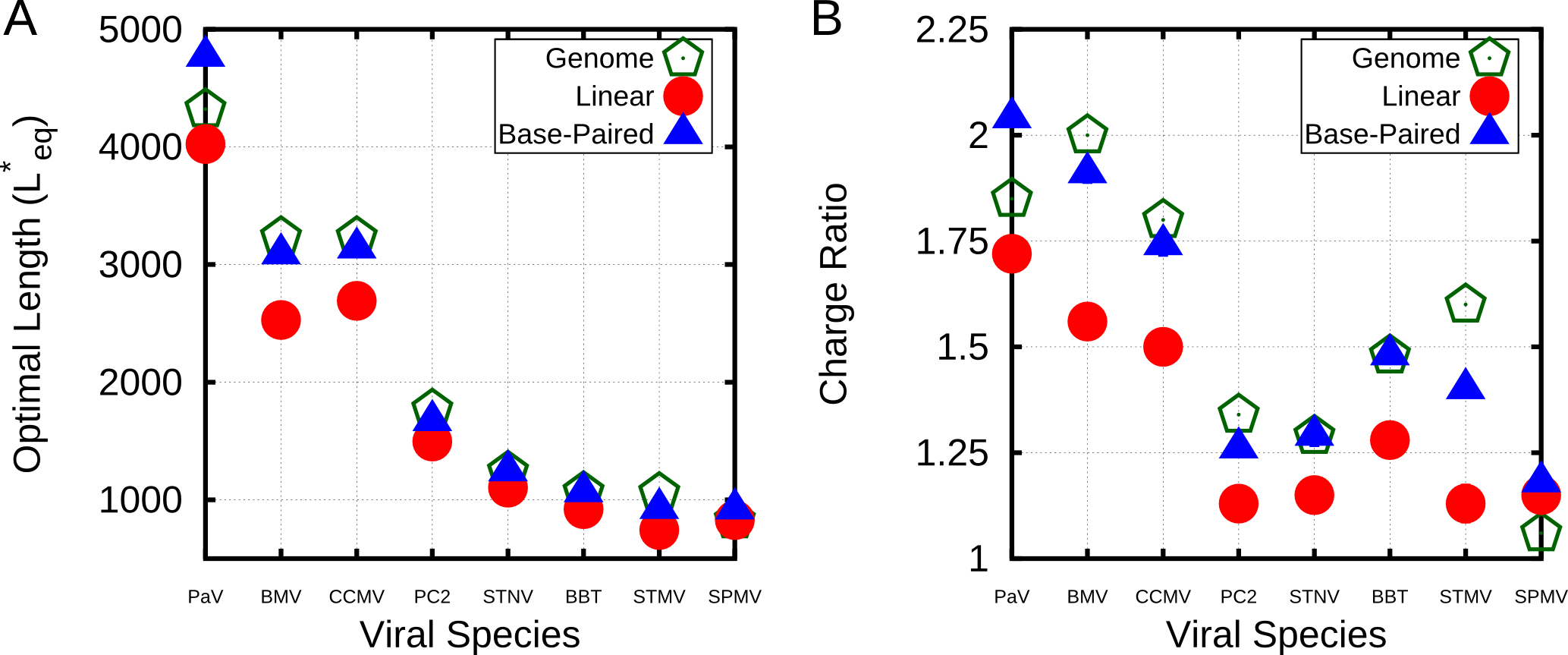}}
\caption{Comparison between viral genome lengths and calculated thermodynamic optimal lengths {\bf (A)} and charge ratios {\bf (B)} for models based on the indicated viral capsid structures. Predicted optimal lengths are shown for linear polyelectrolytes ({\Large\textcolor{red}{$\bullet$}} symbols) and model NAs (\textcolor{blue}{$\blacktriangle$} symbols) with 50\% base-pairing. Viral genome lengths are shown with \textcolor{OliveGreen}{\pentagon} symbols. Error bars fall within the symbol sizes.
\label{fig4}
}
\end{figure}

The predicted values of $\Lopt$ for linear polyelectrolytes and base-paired NAs are compared to the actual viral genome lengths in Fig.~\ref{fig4}. We see that overcharging (charge ratios larger than 1, Fig.~\ref{fig4}B) is predicted for all structures. Furthermore, while the values of $\Lopt$ predicted for linear polyelectrolytes fall short of the viral genome lengths for all investigated structures except for SPMV (whose virion has an unusually low charge ratio), $\Lopt$ for the NA models are relatively close to the viral genome lengths for most structures. Recalling that $\Lopt$ sets an upper bound on length of a polymer that can be efficiently packaged during assembly (Fig.~\ref{fig2}B), this result suggests that the geometric effects of base-pairing contribute to spontaneous packaging of viral genomes. The largest difference between $\Lopt$ and genome length occurs for STMV. This discrepancy may reflect the fact that we used a NA base-pairing fraction of $\fbp=0.5$ whereas 57\% of nucleotides participate in secondary structure elements in the STMV crystal structure \cite{Larson1998,Zeng2012} (lower fractions of nucleotides are resolved in other virion structures, suggesting lower values of $\fbp$).

\begin{table}
\centering
\begin{tabular}{c|c|c|c|c|c}
	Virus	&	Inradius (nm)	&	ARM Length/Net Charge	&	Genome Length	&	$\Lopt$	&	Occupied Volume Fraction\\
\hline
	PaV		&	13.0		&	48/+13	&	4322		&	4766		&	0.074\\
	CCMV		&	11.5		&	48/+10	&	3233		&	3136		&	0.099\\
	BMV		&	11.5		&	44/+9	&	3233		&	3087		&	0.093\\
	PC2		&	8.0		&	43/+22	&	1767		&	1672		&	0.265\\
	STNV		&	7.7		&	28/+16	&	1239		&	1242		&	0.240\\
	BBT		&	7.5		&	27/+12	&	1066		&	1058		&	0.209\\
	STMV		&	7.2		&	19/+11	&	1058		&	922		&	0.232\\
	SPMV		&	6.8		&	20/+13	&	826		&	918		&	0.276
\end{tabular}
\caption{Details for the models of biological capsids studied in this article. The capsid inradius (distance from capsid center to face center), number of residues in the arginine rich motif (ARM), and net charge of the ARM and inner surface are features used to build these models. The viral genome length is then presented for comparison to the value of $\Lopt$ predicted for the base-paired model. Finally, the fraction of occupied volume within the capsid is given for the base-paired model at the optimal length.
\label{tab1}
}
\end{table}

\section{Discussion}

We have shown that assembly simulations and equilibrium calculations based on our coarse-grained model predict optimal NA lengths which are overcharged and relatively close to actual genome sizes for a number of viruses. This finding contrasts with earlier continuum models solved under an assumption of spherical symmetry, which required either a Donnan potential \cite{Ting2011, Ni2012} or irreversible charge renormalization of the NA \cite{Belyi2006} (see SI section~\ref{sec:CounterionCondensation}) to account for overcharging. Our results (Figs.~\ref{fig2}, \ref{fig3}, \ref{fig4}) show that the optimal genome length is determined by a complex interplay between capsid charge, capsid size, excluded volume, and RNA structure.

\begin{figure}
\centerline{\includegraphics[width=0.5\columnwidth]{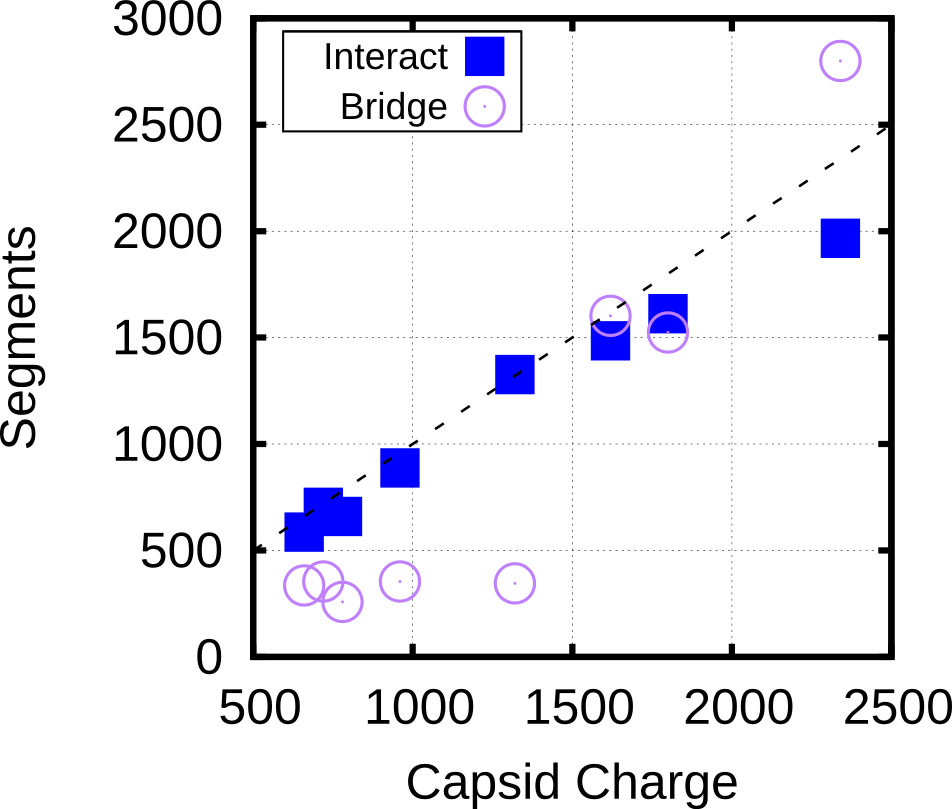}}
\caption{Number of NA segments that directly interact with positively charged ARM segments (interaction energy $\leq -0.5 \kt$, \textcolor{blue}{$\blacksquare$} symbols) and bridging segments (interaction energy $> -0.5 \kt$, {\Large\textcolor{Orchid}{$\circ$}}). The numbers are calculated at the optimal length $\Lopt$ for each capsid shown in Fig.~\ref{fig4} using the base-paired model. For visual reference, the dashed line indicates a 1:1 correspondence between capsid charge and number of nucleotides.
\label{fig5}
}
\end{figure}

{\bf The origins of overcharging.} Analysis of conformations of encapsulated polymers in our simulations shows that overcharging arises as a consequence of geometry and electrostatic screening. The presence of discrete positive charges located in ARMs (or on the capsid surface) combined with nm-scale screening of electrostatics limits the number of direct NA-protein electrostatic interactions; the remaining nucleotides are found in segments which bridge the gaps between positive charges. These interconnecting (bridging) segments are the primary origin of overcharging. Earlier approaches which assumed spherical symmetry could not capture these bridging segments and thus did not predict overcharging. The significance of bridging segments to overcharging is clearly revealed by the dependence of optimal length on capsid size under constant ARM length (Fig.~\ref{fig3}B). For $\inrad \ge 12.5$ nm, the amount of NA interacting with the ARMs is constant, while bridging lengths increase with capsid radius (SI section~\ref{sec:bridging}) due to the increased typical distance between charges on different ARMs. The increase in $\Lopt$ with capsid radius in these calculations can be attributed entirely to increased bridge lengths.

Although the amounts of bridging segments in the biological capsid models depend on many control parameters (e.g. charge, volume, packing fraction, RNA structure), we also confirmed the significance of bridging segments to overcharging in these calculations. Figure ~\ref{fig5} breaks down the $\Lopt$ into the number of segments which interact with positive ARM charges and the number of segments which are bridging. If one counts only the NA segments that directly interact with capsid charges, the resulting charge ratio is slightly less than 1 for each of these capsids. However, when the bridging segments are included, all the capsids are overcharged. Interestingly, more bridging segments are found in the larger $T{=}3$ capsids (56\% bridging) than in $T{=}1$ capsids (25\% bridging), contributing to the larger predicted charge ratios for $T{=}3$ capsids (Fig.~\ref{fig4}B). Though the fraction of nucleotides closely interacting with protein in capsids is difficult to measure experimentally, it might be estimated from the amounts of RNA resolved in crystallographic or EM structures.

We emphasize that our coarse-grained model is designed to incorporate the minimal set of features required to explain the thermodynamic stability of viral particles, and thus neglects some factors that contribute to packaging specific NAs. The \emph{in vivo} experiments in Ref. \cite{Ni2012} on brome mosaic virus (BMV) showed that even charge-conserving mutations to ARM residues could affect the amounts and types of packaged RNA, possibly by interfering with coordination of RNA replication and encapsidation \cite{rao2006}. Similarly, packaging signals, or regions of RNA that have sequence-specific interactions with the capsid protein, are known for some viruses (e.g. HIV \cite{D'Souza2005} or MS2 and satellite tobacco necrosis virus (STNV) \cite{Bunka2011, Borodavka2012}). Packaging signals could be added to our model to investigate how they favor selective assembly around the viral genome through kinetic \cite{ Borodavka2012} or thermodynamic effects. The fact that our model predicts $\Lopt$ for STNV close to its genome length without accounting for sequence specificity may suggest that packaging signals have only a small effect on the thermodynamic $\Lopt$.

In conclusion, our results elucidate the connection between structure and assembly for viral capsids. Firstly, our simulations show that `overcharged' capsids are favored thermodynamically and kinetically, even in the absence of cellular factors or other external effects. Secondly, our results delineate how the genome length which is most favorable for assembly depends on virus-specific quantities such as capsid charge, capsid volume, and genomic tertiary structure. While the correlation between predicted optimal lengths and viral genome sizes suggests that our results have biological relevance, the physical foundations of our model can be tested via controlled \emph{in vitro} experiments. As noted above, several existing experimental observations agree with our results. A positive correlation between protein charge and amounts of packaged RNA has been demonstrated through experiments in which the charge on capsid protein ARMs was altered by mutagenesis (e.g. \cite{dong1998, Kaplan1998,Venter2009}). Competition assays \cite{Porterfield2010, Comas-Garcia2012}, in which two species of NAs or other polymers compete for packaging by a limiting concentration of capsid proteins, offer a quantitative estimate of $\Lopt$. For example, our prediction that $\Lopt$ for CCMV is roughly consistent with the genome length (Fig.~\ref{fig4}) agrees with the observation that CCMV proteins preferentially package longer RNAs, up to the wildtype genome length, over shorter RNAs in competition assays \cite{Comas-Garcia2012}. Now, it is possible to quantitatively test the predictions of our model for the dependence of $\Lopt$ on protein charge and salt concentration through similar competition assays in which NA length preferences are observed for proteins with charge altered by mutagenesis under different ionic strengths. Similarly, our prediction that base-pairing increases $\Lopt$ can be evaluated by comparison of assembly experiments on RNA and synthetic polyelectrolytes (e.g. polystyrene sulfonate) or RNA with base-pairing inhibited through chemical modification (e.g. etheno-RNA \cite{Dhason2012}). Our simulations predict that above the optimal length for a linear polyelectrolyte, only base-paired RNA will be packaged in high yields of well-formed capsids.

\section{Materials and Methods}
\textbf{Model Potentials and Parameters} We have extended a model for empty capsid assembly \cite{Wales2005, Fejer2009, Johnston2010} to describe assembly around NAs. A complete listing of the interaction potentials is provided in SI section~\ref{sec:S1A}; we summarize them here. The pseudoatoms in the capsid subunit model are illustrated in Fig.~\ref{fig1}. Subunit assembly is mediated through an attractive Morse potential between Attractor (`A') pseudoatoms located at each subunit vertex. The Top (`T') pseudoatoms interact with other `T' psuedoatoms through a potential consisting of the repulsive term of the LJ potential, the radius of which is chosen to favor a subunit-subunit angle consistent with a dodecahedron (116 degrees). The Bottom (`B') pseudoatom has a repulsive LJ interaction with `T' pseudoatoms, intended to prevent `upside-down' assembly. The `T', `B', and `A' pseudoatoms form a rigid body \cite{Wales2005, Fejer2009, Johnston2010}.

To model electrostatic interaction with a negatively charged NA or polyelectrolyte we extend the model as follows. Firstly, to better represent the capsid shell we add a layer of `Excluder' pseudoatoms which have a repulsive LJ interaction with the polyelectrolyte and the ARMs. Each ARM is modeled as a bead-spring polymer, with one bead per amino acid. The `Excluders' and first ARM segment are part of the subunit rigid body. ARM beads interact through repulsive Lennard-Jones interactions and, if charged, electrostatic interactions modeled by a Debye-Huckel potential. Comparison to Coulomb interactions with explicit counterions is shown in Fig.~\ref{fig3}d and in SI section~\ref{sec:salt}. We also show that the results do not change significantly when experimentally relevant concentrations of divalent cations are added to the system (Fig. \ref{figS1}). The ability of the Debye-Huckel model to provide a reasonable representation of electrostatics in the system can be understood based on the relatively low packing fractions (see Table~\ref{tab1}) within the assembled capsids and the fact that the relevant experimental and physiological conditions correspond to moderate to high salt concentrations.

\textbf{Simulation Methods} Simulations were performed with the Brownian Dynamics algorithm of HOOMD, which uses the Langevin equation to evolve positions and rigid body orientations in time \cite{Anderson2008, nguyen2011,LeBard2012}. Simulations were run using a set of fundamental units. The fundamental energy unit is selected to be $\eunit\equiv1 \kt$. The unit of length $\dunit$ is set to the circumradius of a pentagonal subunit, which is taken to be $1\dunit\equiv5$ nm so that the dodecahedron inradius of $1.46\dunit=7.3$ nm gives an interior volume consistent with that of the smallest $T{=}1$ capsids (SI section ~\ref{sec:specificCapsids}). Assembly simulations were run at least 10 times for each set of parameters, each of which were concluded at $2\times10^8$ time steps. The following parameters values were used in all of our dynamical assembly simulations: $\ld=1$ nm, $\mbox{box size} = 200 \times 200 \times 200$ nm, $\mbox{subunit concentration}=12 \mu$M. During calculation of the thermodynamic optimal polymer length $\Lopt$, calculations were run at least $1\times10^7$ timesteps, with equilibrium assessed after convergence. Standard error was obtained based on averages of multiple ($\ge3$) independent simulations. Separate calculations of $\Lopt$ were also performed using using the Widom test-particle method \cite{Widom1963} as extended to calculate polymer residual chemical potentials \cite{Kumar1991, Elrad2010} (described in more detail in SI section~\ref{sec:equilibrium}). Snapshots from simulations were visualized using VMD \cite{Humphrey1996}.

\section{acknowledgments}
We gratefully acknowledge Chuck Knobler, William Gelbart, and Adam Zlotnick for insightful discussions and critical reads of the manuscript. This work was supported by Award Number R01AI080791 from the National Institute Of Allergy And Infectious Diseases. Computational resources were provided by the NSF through XSEDE computing resources (Longhorn, Condor, \& Keeneland) and the Brandeis HPCC.

\bibliographystyle{plain}
\bibliography{Perlmutter2013.bib}

\begin{thebibliography}{10}

\bibitem{webnote}
Movies are available at
  http://www.brandeis.edu/departments/physics/hagan/movies/index.html.

\bibitem{Anderson2008}
Joshua~A. Anderson, Chris~D. Lorenz, and A.~Travesset.
\newblock General purpose molecular dynamics simulations fully implemented on
  graphics processing units.
\newblock {\em J. Comput. Phys.}, 227(10):5342--5359, 2008.

\bibitem{Angelescu2006}
D.~G. Angelescu, R.~Bruinsma, and P.~Linse.
\newblock {M}onte {C}arlo simulations of polyelectrolytes inside viral capsids.
\newblock {\em Phys. Rev. E}, 73(4):041921, 2006.

\bibitem{Ban1995}
N.~Ban and A.~McPherson.
\newblock The structure of satellite panicum mosaic virus at 1.9 {\aa}
  resolution.
\newblock {\em Nature Structural \& Molecular Biology}, 2(10):882--890, 1995.

\bibitem{Beglov1994}
D.~Beglov and B.~Roux.
\newblock Finite representation of an infinite bulk system -- solvent boundary
  potential for computer simulations.
\newblock {\em J. Chem. Phys.}, 100(12):9050--9063, 1994.

\bibitem{Belyi2006}
V.~A. Belyi and M.~Muthukumar.
\newblock Electrostatic origin of the genome packing in viruses.
\newblock {\em Proc. Natl. Acad. Sci. U. S. A.}, 103(46):17174--17178, 2006.

\bibitem{Bentley1987}
GA~Bentley, A.~Lewit-Bentley, L.~Liljas, U.~Skoglund, M.~Roth, and T.~Unge.
\newblock Structure of rna in satellite tobacco necrosis virus: A low
  resolution neutron diffraction study using 1h2o2h2o solvent contrast
  variation.
\newblock {\em J. Mol. Biol.}, 194(1):129--141, 1987.

\bibitem{Borodavka2012}
Alexander Borodavka, Roman Tuma, and Peter~G. Stockley.
\newblock Evidence that viral {RNA}s have evolved for efficient, two-stage
  packaging.
\newblock {\em Proc. Natl. Acad. Sci. U. S. A.}, 109(39):15769--15774, 2012.

\bibitem{Bunka2011}
D.~H.~J. Bunka, S.~W. Lane, C.~L. Lane, E.~C. Dykeman, R.~J. Ford, A.~M.
  Barker, R.~Twarock, S.~E.~V. Phillips, and P.~G. Stockley.
\newblock Degenerate {RNA} packaging signals in the genome of satellite tobacco
  necrosis virus: Implications for the assembly of a {T=1} capsid.
\newblock {\em J. Mol. Biol.}, 413(1):51--65, 2011.

\bibitem{Cadena-Nava2012}
Ruben~D. Cadena-Nava, Mauricio Comas-Garcia, Rees~F. Garmann, A.~L.~N. Rao,
  Charles~M. Knobler, and William~M. Gelbart.
\newblock Self-assembly of viral capsid protein and {RNA} molecules of
  different sizes: Requirement for a specific high protein/{RNA} mass ratio.
\newblock {\em J. Virol.}, 86(6):3318--3326, 2012.

\bibitem{Carrillo-Tripp2009}
M.~Carrillo-Tripp, C.M. Shepherd, I.A. Borelli, S.~Venkataraman, G.~Lander,
  P.~Natarajan, J.E. Johnson, C.L. Brooks~III, and V.S. Reddy.
\newblock Viperdb2: an enhanced and web api enabled relational database for
  structural virology.
\newblock {\em Nucleic Acids Res.}, 37(suppl 1):D436--D442, 2009.

\bibitem{Caspar1962}
D.~L.~D. Caspar and A.~Klug.
\newblock Physical principles in construction of regular viruses.
\newblock {\em Cold Spring Harbor Symp. Quant. Biol.}, 27:1--24, 1962.

\bibitem{Comas-Garcia2012}
Mauricio Comas-Garcia, Ruben~D. Cadena-Nava, A.~L.~N. Rao, Charles~M. Knobler,
  and William~M. Gelbart.
\newblock In vitro quantification of the relative packaging efficiencies of
  single-stranded {RNA} molecules by viral capsid protein.
\newblock {\em J. Virol.}, 2012.

\bibitem{Dhason2012}
Mary~S. Dhason, Joseph C.~Y. Wang, Michael~F. Hagan, and Adam Zlotnick.
\newblock Differential assembly of hepatitis {B} virus core protein on single-
  and double-stranded nucleic acid suggest the ds{DNA}-filled core is
  spring-loaded.
\newblock {\em Virology}, 430(1):20--29, 2012.

\bibitem{dong1998}
X.~F. Dong, P.~Natarajan, M.~Tihova, J.~E. Johnson, and A.~Schneemann.
\newblock Particle polymorphism caused by deletion of a peptide molecular
  switch in a quasiequivalent icosahedral virus.
\newblock {\em J. Virol.}, 72(7):6024--6033, 1998.

\bibitem{D'Souza2005}
V.~D'Souza and M.~F. Summers.
\newblock How retroviruses select their genomes.
\newblock {\em Nat. Rev. Microbiol.}, 3(8):643--655, 2005.

\bibitem{Elrad2010}
O.M. Elrad and M.~F. Hagan.
\newblock Encapsulation of a polymer by an icosahedral virus.
\newblock {\em Phys. Biol.}, 7:045003, 2010.

\bibitem{ElSawy2011}
Karim~M. ElSawy, Leo S.~D. Caves, and Reidun Twarock.
\newblock On the origin of order in the genome organization of ss{RNA} viruses.
\newblock {\em Biophys. J.}, 101(4):774--780, 2011.

\bibitem{Fejer2009}
Szilard~N. Fejer, Tim~R. James, Javier Hernandez-Rojas, and David~J. Wales.
\newblock Energy landscapes for shells assembled from pentagonal and hexagonal
  pyramids.
\newblock {\em Phys. Chem. Chem. Phys.}, 11(12):2098--2104, 2009.

\bibitem{Gopal2012}
A.~Gopal, Z.H. Zhou, C.M. Knobler, and W.M. Gelbart.
\newblock Visualizing large rna molecules in solution.
\newblock {\em RNA}, 18(2):284--299, 2012.

\bibitem{Hagan2006}
M.~F. Hagan and D.~Chandler.
\newblock Dynamic pathways for viral capsid assembly.
\newblock {\em Biophys. J.}, 91(1):42--54, 2006.

\bibitem{Hagan2011}
M.~F. Hagan, O.~M. Elrad, and R.~L. Jack.
\newblock Mechanisms of kinetic trapping in self-assembly and phase
  transformation.
\newblock {\em J. Chem. Phys.}, 135:104115, 2011.

\bibitem{harding1991}
R.M. Harding, T.M. Burns, J.L. Dale, et~al.
\newblock Virus-like particles associated with banana bunchy top disease
  contain small single-stranded dna.
\newblock {\em J. Gen. Virol.}, 72(2):225--230, 1991.

\bibitem{Hu2008a}
T.~Hu, R.~Zhang, and B.~I. Shklovskii.
\newblock Electrostatic theory of viral self-assembly.
\newblock {\em Physica A}, 387(12):3059--3064, 2008.

\bibitem{Humphrey1996}
W.~Humphrey, A.~Dalke, and K.~Schulten.
\newblock {VMD}: Visual molecular dynamics.
\newblock {\em J. Mol. Graph.}, 14(1):33--38, 1996.

\bibitem{Johnston2010}
I.~G. Johnston, A.~A. Louis, and J.~P.~K. Doye.
\newblock Modelling the self-assembly of virus capsids.
\newblock {\em J. Phys.: Condens. Matter}, 22(10), 2010.

\bibitem{Jones1984}
T.A. Jones and L.~Liljas.
\newblock Structure of satellite tobacco necrosis virus after crystallographic
  refinement at 2.5 {\aa} resolution.
\newblock {\em J. Mol. Biol.}, 177(4):735--767, 1984.

\bibitem{Kaplan1998}
I.~B. Kaplan, L.~Zhang, and P.~Palukaitis.
\newblock Characterization of cucumber mosaic virus - v. cell-to-cell movement
  requires capsid protein but not virions.
\newblock {\em Virology}, 246(2):221--231, 1998.

\bibitem{Khayat2011}
R.~Khayat, N.~Brunn, J.A. Speir, J.M. Hardham, R.G. Ankenbauer, A.~Schneemann,
  and J.E. Johnson.
\newblock The 2.3-angstrom structure of porcine circovirus 2.
\newblock {\em J. Virol.}, 85(15):7856--7862, 2011.

\bibitem{Kivenson2010}
A~Kivenson and MF~Hagan.
\newblock Mechanisms of viral capsid assembly around a polymer.
\newblock {\em Biophys. J.}, Vol 99 Issue 2(2):619--628, 2010.

\bibitem{Kler2012}
Stanislav Kler, Roi Asor, Chenglei Li, Avi Ginsburg, Daniel Harries, Ariella
  Oppenheim, Adam Zlotnick, and Uri Raviv.
\newblock {RNA} encapsidation by {SV40}-derived nanoparticles follows a rapid
  two-state mechanism.
\newblock {\em J. Am. Chem. Soc.}, 134(21):8823--8830, 2012.

\bibitem{Kumar1991}
SK~Kumar, I~Szleifer, and AZ~Panagiotopoulos.
\newblock Determination of the chemical-potentials of polymeric systems from
  monte-carlo simulations.
\newblock {\em Phys. Rev. Lett.}, 66(22):2935--2938, JUN 3 1991.

\bibitem{Larson1998}
Steven~B. Larson, John Day, Aaron Greenwood, and Alexander McPherson.
\newblock Refined structure of satellite tobacco mosaic virus at 1.8 Å
  resolution.
\newblock {\em J. Mol. Biol.}, 277(1):37--59, 1998.

\bibitem{LeBard2012}
David~N. LeBard, Benjamin~G. Levine, Philipp Mertmann, Stephen~A. Barr, Arben
  Jusufi, Samantha Sanders, Michael~L. Klein, and Athanassios~Z.
  Panagiotopoulos.
\newblock Self-assembly of coarse-grained ionic surfactants accelerated by
  graphics processing units.
\newblock {\em Soft Matter}, 8(8):2385--2397, 2012.

\bibitem{li2002}
P.~P. Li, A.~Nakanishi, S.~W. Clark, and H.~Kasamatsu.
\newblock Formation of transitory intrachain and interchain disulfide bonds
  accompanies the folding and oligomerization of simian virus 40 vp1 in the
  cytoplasm.
\newblock {\em Proc. Natl. Acad. Sci. U. S. A.}, 99(3):1353--1358, 2002.

\bibitem{LosdorferBozic2012}
Anze Losdorfer~Bozic, Antonio Siber, and Rudolf Podgornik.
\newblock How simple can a model of an empty viral capsid be? charge
  distributions in viral capsids.
\newblock {\em Journal of Biological Physics}, 38(4):657--671, 2012.

\bibitem{Lucas2002}
R.~W. Lucas, S.~B. Larson, and A.~McPherson.
\newblock The crystallographic structure of brome mosaic virus.
\newblock {\em J. Mol. Biol.}, 317(1):95--108, 2002.

\bibitem{MacKerell1998}
A.~D.~Jr. MacKerell, D.~Bashford, M.~Bellott, R.~L.~Jr. Dunbrack, J.~D.
  Evaseck, M.~J. Field, S.~Fischer, J.~Gao, H.~Guo, S.~Ha, D.~Joseph-McCarthy,
  L.~Kuchnir, K.~kuczera, F.~T.~K. Lau, C.~Mattos, S.~Michnick, T.~Ngo, D.~T.
  Nguyen, B.~Prodhom, W.~E.~III Reiher, B.~Roux, M.~Schlenkrich, J.~C. Smith,
  R.~Stote, J.~Straub, M.~Watanabe, J.~Wiorkiewicz-Kuczera, D.~Yin, and Martin
  Karplus.
\newblock All-atom empirical potential for molecular modeling and dynamics
  studies of proteins.
\newblock {\em J. Phys. Chem. B}, 102(18):3586--3616, 1998.

\bibitem{Mackerell2004}
Alexander~D. Mackerell.
\newblock Empirical force fields for biological macromolecules: Overview and
  issues.
\newblock {\em J. Comput. Chem.}, 25(13):1584--1604, 2004.

\bibitem{Mahalik2012}
J.~P. Mahalik and M.~Muthukumar.
\newblock Langevin dynamics simulation of polymer-assisted virus-like assembly.
\newblock {\em J. Chem. Phys.}, 136(13):135101, 2012.

\bibitem{nguyen2011}
Trung~Dac Nguyen, Carolyn~L. Phillips, Joshua~A. Anderson, and Sharon~C.
  Glotzer.
\newblock Rigid body constraints realized in massively-parallel molecular
  dynamics on graphics processing units.
\newblock {\em Comput. Phys. Commun.}, 182(11):2307--2313, 2011.

\bibitem{Ni2012}
P.~Ni, Z.~Wang, X.~Ma, N.~C. Das, P.~Sokol, W.~Chiu, B.~Dragnea, M.~F. Hagan,
  and C.~C Kao.
\newblock An examination of the electrostatic interactions between the
  n-terminal tail of the coat protein and {RNA} in brome mosaic virus.
\newblock {\em J. Mol. Biol.}, 419:284--300, 2012.

\bibitem{Porterfield2010}
J.~Zachary Porterfield, Mary~Savari Dhason, Daniel~D. Loeb, Michael Nassal,
  Stephen~J. Stray, and Adam Zlotnick.
\newblock Full-length hepatitis {B} virus core protein packages viral and
  heterologous {RNA} with similarly high levels of cooperativity.
\newblock {\em J. Virol.}, 84(14):7174--7184, 2010.

\bibitem{rao2006}
A.~L.~N. Rao.
\newblock Genome packaging by spherical plant {RNA} viruses.
\newblock In {\em Annual Review of Phytopathology}, volume~44 of {\em Annual
  Review of Phytopathology}, pages 61--87. 2006.

\bibitem{Schneemann2006}
A.~Schneemann.
\newblock The structural and functional role of {RNA} in icosahedral virus
  assembly.
\newblock {\em Annu. Rev. Microbiol.}, 60:51--67, 2006.

\bibitem{Serwer1999}
Philip Serwer and Gary~A. Griess.
\newblock Advances in the separation of bacteriophages and related particles.
\newblock {\em J. Chromatogr. B}, 722(1–2):179--190, 1999.

\bibitem{Serwer1995}
Philip Serwer, Saeed~A. Khan, and Gary~A. Griess.
\newblock Non-denaturing gel electrophoresis of biological nanoparticles:
  viruses.
\newblock {\em J. Chromatogr. A}, 698(1–2):251--261, 1995.

\bibitem{siber2008}
A.~Siber and R.~Podgornik.
\newblock Nonspecific interactions in spontaneous assembly of empty versus
  functional single-stranded {RNA} viruses.
\newblock {\em Phys. Rev. E}, 78:051915, 2008.

\bibitem{Speir1995}
J.~A. Speir, S.~Munshi, G.~J. Wang, T.~S. Baker, and J.~E. Johnson.
\newblock Structures of the native and swollen forms of cowpea chlorotic mottle
  virus determined by x-ray crystallography and cryoelectron microscopy.
\newblock {\em Structure}, 3(1):63--78, 1995.

\bibitem{Speir2012}
Jeffrey~A. Speir and John~E. Johnson.
\newblock Nucleic acid packaging in viruses.
\newblock {\em Curr. Opin. Struct. Biol.}, 22(1):65--71, 2012.

\bibitem{Stehle1996}
T.~Stehle, S.~J. Gamblin, Y.~W. Yan, and S.~C. Harrison.
\newblock The structure of simian virus 40 refined at 3.1 angstrom resolution.
\newblock {\em Structure}, 4(2):165--182, 1996.

\bibitem{sugita1999}
Y.~Sugita and Y.~Okamoto.
\newblock Replica-exchange molecular dynamics method for protein folding.
\newblock {\em Chem. Phys. Lett.}, 314(1):141--151, 1999.

\bibitem{Tang2001}
L.~Tang, K.N. Johnson, L.A. Ball, T.~Lin, M.~Yeager, and J.E. Johnson.
\newblock The structure of pariacoto virus reveals a dodecahedral cage of
  duplex rna.
\newblock {\em Nature Structural \& Molecular Biology}, 8(1):77--83, 2001.

\bibitem{Ting2011}
Christina~L. Ting, Jianzhong Wu, and Zhen-Gang Wang.
\newblock Thermodynamic basis for the genome to capsid charge relationship in
  viral encapsidation.
\newblock {\em Proc. Natl. Acad. Sci. U. S. A.}, 108(41):16986--16991, 2011.

\bibitem{Valegard1997}
K.~Valegard, J.~B. Murray, N.~J. Stonehouse, S.~vandenWorm, P.~G. Stockley, and
  L.~Liljas.
\newblock The three-dimensional structures of two complexes between recombinant
  {MS2} capsids and {RNA} operator fragments reveal sequence-specific
  protein-{RNA} interactions.
\newblock {\em J. Mol. Biol.}, 270(5):724--738, 1997.

\bibitem{Schoot2005}
P.~van~der Schoot and R.~Bruinsma.
\newblock Electrostatics and the assembly of an {RNA} virus.
\newblock {\em Phys. Rev. E}, 71(6):061928, 2005.

\bibitem{Venter2009}
P.~A. Venter, D.~Marshall, and A.~Schneemann.
\newblock Dual roles for an arginine-rich motif in specific genome recognition
  and localization of viral coat protein to {RNA} replication sites in flock
  house virus-infected cells.
\newblock {\em J. Virol.}, 83(7):2872--2882, 2009.

\bibitem{Wales2005}
D.~J. Wales.
\newblock The energy landscape as a unifying theme in molecular science.
\newblock {\em Phil. Trans. R. Soc. A}, 363(1827):357--375, 2005.

\bibitem{Widom1963}
B.~Widom.
\newblock Some topics in the theory of fluids.
\newblock {\em J. Chem. Phys.}, 39(11):2808--2812, 1963.

\bibitem{Yoffe2008}
A.~M. Yoffe, P.~Prinsen, A.~Gopal, C.~M. Knobler, W.~M. Gelbart, and
  A.~Ben-Shaul.
\newblock Predicting the sizes of large {RNA} molecules.
\newblock {\em Proc. Natl. Acad. Sci. U. S. A.}, 105(42):16153--16158, 2008.

\bibitem{Zeng2012}
Y.~Zeng, S.B. Larson, C.E. Heitsch, A.~McPherson, and S.C. Harvey.
\newblock A model for the structure of satellite tobacco mosaic virus.
\newblock {\em J. Struct. Biol.}, 2012.

\bibitem{Zhang2004a}
Z.~L. Zhang and S.~C. Glotzer.
\newblock Self-assembly of patchy particles.
\newblock {\em Nano Lett.}, 4(8):1407--1413, 2004.

\bibitem{Zlotnick2013}
Adam Zlotnick, J.~Zachary Porterfield, and Joseph Che-Yen Wang.
\newblock To build a virus on a nucleic acid substrate.
\newblock {\em Biophys. J.}, 104(7):1595--1604, 2013.

\end{thebibliography}

\pagebreak 

\setcounter{equation}{0}
\renewcommand{\theequation}{S\arabic{equation}}
\setcounter{figure}{0}
\renewcommand{\thefigure}{S\arabic{figure}}
\setcounter{table}{0}
\renewcommand{\thetable}{S\arabic{table}}
\setcounter{section}{0}
\renewcommand{\thesection}{S\arabic{section}}

\section{Additional results}
\label{sec:additionalResults}

\subsection{Effect of salt concentration}
\label{sec:salt}

To evaluate the effect of the approximations made in the Debye-Huckel treatment of electrostatics, we performed a number of simulations using the primitive model representation of electrostatics and explicit ions to represent neutralizing counterions and added salt. Ions are modeled as repulsive spheres (Eq.~\ref{eq:LJ} below) and electrostatics are calculated according to Coulomb interactions (Eq.~\ref{eq:coulomb} below) with the relative permittivity set to 80.

\begin{figure}
\centering{\includegraphics[width=0.99\columnwidth]{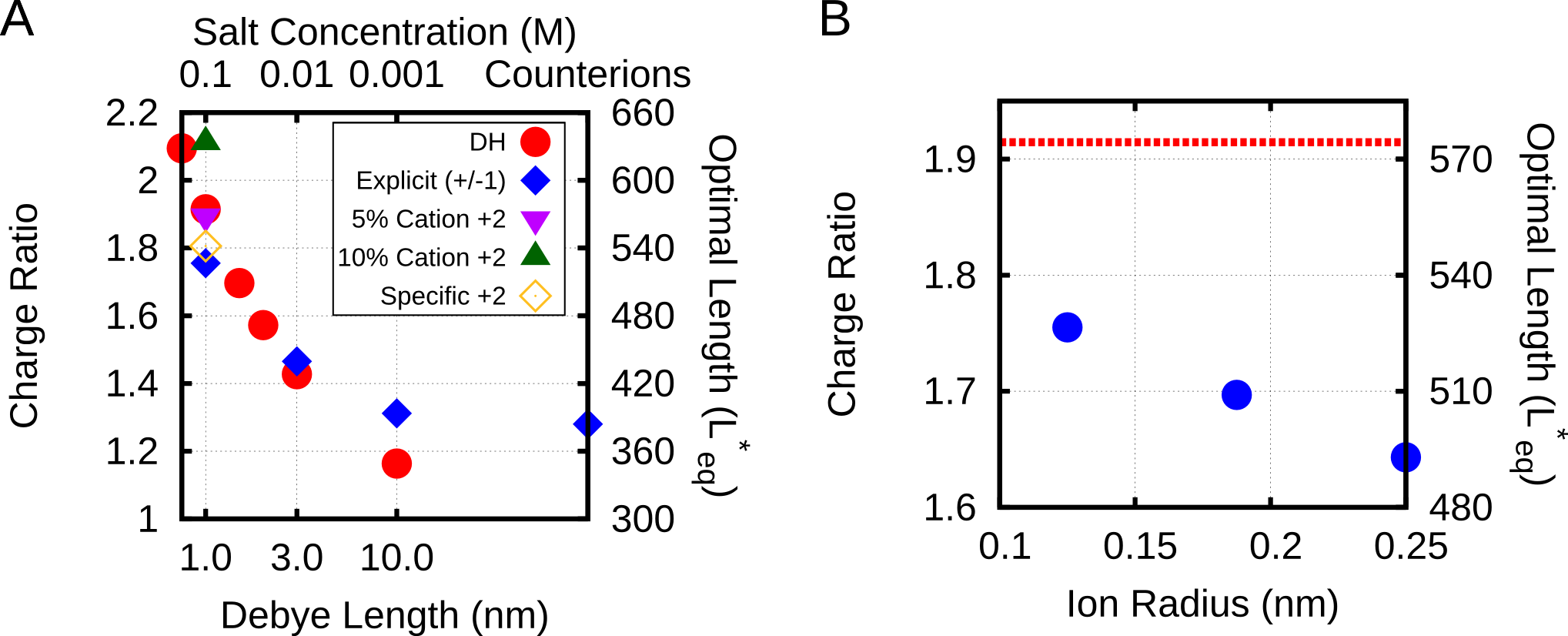}}
\caption{{\bf (A)}Effect of ionic strength and comparison between Debye-Huckel interactions and explicit ions. The thermodynamic optimum lengths $\Lopt$ and corresponding optimal charge ratios are shown as functions of the Debye screening length (bottom axis) and ionic strength (top axis), calculated with simulations using Debye-Huckel (DH) interactions ({\Large\textcolor{red}{$\bullet$}} symbols) or Coulomb interactions with explicit ions,  1:1 salt and no divalent cations (\textcolor{blue}{\DiamondSolid} symbols), 5\% 2:1 salt (\textcolor{purple}{\TriangleDown} symbols) or 10\% 2:1 salt (\textcolor{green}{\TriangleUp} symbols). An additional system with monovalent free ions and divalent cations irreversibly bound to the polyelectrolyte is also presented ({\Large\textcolor{orange}{$\diamond$}} symbols). Calculations were performed using the simple capsid model (Fig.~\ref{fig1} in the main text) and a linear polyelectrolyte. All ion radii were set to 0.125 nm. {\bf (B)}Effect of varying the ion radius ({\Large\textcolor{blue}{$\bullet$}} symbols) at 100 mM monovalent ions, compared with the DH result (dashed red line).
\label{figS1}
}
\end{figure}

As shown in Fig.~\ref{figS1}A, the optimal length $\Lopt$ and charge ratio increase with increasing ionic strength (i.e. decreasing Debye length $\ld$). This effect can be explained by the fact that a smaller fraction of NA charges interact with positive capsid charges as the screening length decreases (see the Discussion). Importantly though, the simulations predict overcharging at all salt concentrations investigated ($1\mbox{ mM} \le \Csalt \le 100\mbox{ mM}$). We also present the results of the limiting case where only neutralizing counterions are used (resulting in $\sim 1$ mM cations and 0 anions, for a total ionic strength of $\sim 0.5$ mM). We focus on $\Csalt=100$ mM for all other results in this article.

Over a range of salt concentrations $1\mbox{ mM }< \Csalt < 100\mbox{ mM}$ with 1:1 electrolyte, we see that optimal lengths predicted by simulations using explicit ions or Debye-Huckel interactions agree to within about 10\% (Fig.~\ref{figS1}). The Debye-Huckel simulations slightly overpredict the optimal length at high salt concentrations because they neglect counterion excluded volume, while they underpredict the optimal length at low ionic strength because they neglect ion-ion correlations. As shown in Fig.~\ref{figS1}B, the explicit ion results approach those of the Debye-Huckel model simulations at physiological salt concentrations (100 mM) as the explicit ion radius is decreased, since ion excluded volume is reduced, with the two methods agreeing to within 10\% for the most realistic ion radius (0.125 nm). In all other simulations ionic radii were set to 0.125 nm (i.e. $\sigma=0.25$ nm in Eq.~\ref{eq:LJ} below), which is roughly equal to the radii of Na$^+$ and CL$^-$ ions in the CHARMM force field \cite{Beglov1994, MacKerell1998, Mackerell2004}.

{\bf Effect of divalent cations.} We find that introducing divalent cations increases the optimal length (Fig.~\ref{figS1}A). This result could be anticipated, since the divalent cations preferentially bind to the NA, screening electrostatic repulsions between encapsulated nucleotides. However, the quantitative effect of divalent cations on the predicted optimal length is small at physiologically relevant concentrations (about 1 mM divalent cations). As shown in Fig. \ref{figS1}A, with larger concentrations of divalent cations (about 5 mM and 10 mM divalent) the optimal length increases by 8 and 20\% respectively. Finally, specific binding between Mg$^{2+}$ ions and RNA is known to affect RNA structure. To test the effect of such stably bound divalent cations on optimal length, we constructed a polyelectrolyte with a divalent cation irreversible bound (through a harmonic potential) to every 100th NA segment, in a solution containing 100 mM 1:1 salt. While this model does not capture the structural effects of specific Mg$^{2+}$ binding to RNA, it does represent the fact that these bound cations effectively cancel some NA charges.

\subsection{Semiflexible Polymer}
\begin{figure}
\centering{\includegraphics[width=0.5\columnwidth]{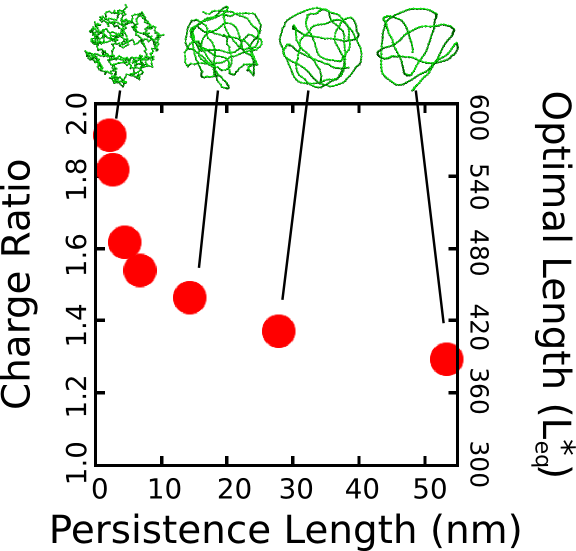}}
\caption{Effect of stiffness on linear polymer equilibrium encapsidation. The thermodynamic optimum lengths $\Lopt$ and charge ratios are shown as a function of persistence length for a linear, semiflexible polyelectrolyte. The simulations were performed using the simple capsid model (Fig.~\ref{fig1} in the main text, with dodecahedron inradius $\inrad=7.3$, and an ARM length of 5 positive charges). Representative snapshots of the encapsidated polymer (taken from simulations at the optimal length) are shown for several persistence lengths. The capsid and ARMs are rendered invisible in these snapshots to enable visibility of the polyelectrolyte.}
\label{figS2}
\end{figure}

Fig.~\ref{figS2} shows $\Lopt$ as a function of polyelectrolyte flexibility, over the range of persistence lengths ($\lp$) relevant to biological nucleic acids. We see that the optimal charge ratio monotonically decreases with increasing persistence length. While this observation could be anticipated on intuitive grounds, the quantitative decrease is substantial --- a 32 \% decrease in optimal charge ratio between our most flexible polymer ($\lp=2.1$ nm) and our stiffest polymer ($\lp=53.4$ nm). The persistence length is obtained by simulating the polymer unencapsidated in solution and fitting the segmental autocorrelation function to an exponential decay, where the persistence length is the decay constant.

\section{Additional model details}
\label{sec:S1}
\subsection{Model potentials and parameters}
\label{sec:S1A}
In our model, all potentials can be decomposed into pairwise interactions. Potentials involving capsomer subunits further decompose into pairwise interactions between their constituent building blocks -- the excluders, attractors, `Top' and `Bottom', and ARM pseudoatoms. It is convenient to write the total energy of the system as the sum of 6 terms: a capsomer-capsomer $U\sub{cc}{}$ part (which does not include interactions between ARM pseudoatoms), capsomer-polymer $U\sub{cp}{}$, capsomer-ARM $U\sub{ca}{}$, polymer-polymer $U\sub{pp}{}$, polymer-ARM $U\sub{pa}{}$, and ARM-ARM $U\sub{aa}{}$ parts. Each is summed over all pairs of the appropriate type:
\begin{align}
U = & \sum_{\mathrm{cap\ }{i}} \sum_{\mathrm{cap\ }{j < i}} U\sub{cc}{}
  + \sum_{\mathrm{cap\ }{i}} \sum_{\mathrm{poly\ }{j}} U\sub{cp}{}
  + \sum_{\mathrm{cap\ }{i}} \sum_{\mathrm{ARM\ }{j}} U\sub{ca}{} + \nonumber \\
  & \sum_{\mathrm{poly\ }{i}} \sum_{\mathrm{poly\ }{j < i}} U\sub{pp}{}
  + \sum_{\mathrm{poly\ }{i}} \sum_{\mathrm{ARM\ }{j}} U\sub{pa}{}
  + \sum_{\mathrm{tail\ }{i}} \sum_{\mathrm{ARM\ }{j < i}} U\sub{aa}{}
\end{align}
where $\sum_{\mathrm{cap\ }{i}} \sum_{\mathrm{cap\ }{j < i}}$ is the sum over all distinct pairs of capsomers in the system, $\sum_{\mathrm{cap\ }{i}} \sum_{\mathrm{poly\ }{j}}$ is the sum over all capsomer-polymer pairs, etc.

The capsomer-capsomer potential $U\sub{cc}{}$ is the sum of the attractive interactions between complementary attractors, and geometry guiding repulsive interactions between `Top' - `Top' pairs and `Top' - `Bottom' pairs. There are no interactions between members of the same rigid body, but ARMs are not rigid and thus there are intra-subunit ARM-ARM interactions. Thus, for notational clarity, we index rigid bodies and non-rigid pseudoatoms in Roman, while the pseudoatoms comprising a particular rigid body are indexed in Greek. E.g., for capsomer $i$ we denote its attractor positions as $\{\bm{a}_{i\alpha}\}$ with the set comprising all attractors $\alpha$, its `Top' positions $\{\bm t_{i\alpha}\}$, and its `Bottom' positions $\{\bm b_{i\alpha}\}$. The capsomer-capsomer interaction potential between two capsomers $i$ and $j$ is then defined as:

\begin{align}
\label{Ucc}
U\sub{cc}{}&(\{\bm a_{i\alpha}\}, \{\bm t_{i\alpha}\}, \{\bm b_{i\alpha}\} , \{\bm a_{j\beta}, \{\bm t_{j\beta}\}, \{\bm b_{j\beta}\})  = \nonumber \\
&\sum_{\alpha,\beta}^{N\sub{t}{}} \varepsilon \LJ{} \left(
\left|\bm{t}_{i\alpha} - \bm{t}_{j\beta} \right|,
\ \sigma\sub{t}{} \right) + \nonumber \\
&\sum_{\alpha,\beta}^{N\sub{b}{},N\sub{t}{}} \varepsilon \LJ{} \left(
\left| \bm{b}_{i\alpha} - \bm{t}_{j\beta} \right|,
\ \sigma\sub{b}{} \right) + \nonumber \\
&\sum_{\alpha,\beta}^{N\sub{a}{}} \varepsilon \Morse{} \left(
\left|\bm{a}_{i\alpha} - \bm{a}_{j\beta} \right|,
\ r\sub{0}{}, \varrho, \rcut \right)
\end{align}
where $\varepsilon$ is an adjustable parameter which both sets the strength of the capsomer-capsomer attraction at each attractor site and scales the repulsive interactions which enforce the dodecahedral geometry. $N\sub{t}{}$, $N\sub{b}{}$, and $N\sub{a}{}$ are the number of `Top', `Bottom', and attractor pseudoatoms respectively in one capsomer, $\sigma\sub{t}{}$ and $\sigma\sub{b}{}$ are the effective diameters of the `Top' -- `Top' interaction and `Bottom' -- `Top' interactions, which are set to 10.5 nm and 9.0 nm respectively throughout this work, $r\sub{0}{}$ is the minimum energy attractor distance, set to 1 nm, $\varrho$ is a parameter determining the width of the attractive interaction, set to 2.5, and $\rcut$ is the cutoff distance for the attractor potential, set to 10.0 nm.

The function $\LJ{}$ is defined as the repulsive component of the Lennard-Jones potential shifted to zero at the interaction diameter:
\begin{equation}
\LJ{}(x,\sigma) \equiv
\left\{  \begin{array}{ll}
	\left(\frac{\sigma}{x}\right)^{12} -1 & : x < \sigma \\
	0 & : \mathrm{otherwise}
\end{array} \right.
\label{eq:LJ}
\end{equation}
The function $\Morse{}$ is a Morse potential:
\begin{equation}
\Morse{}(x,r\sub{0}{},\varrho) \equiv
\left\{  \begin{array}{ll}
\left(e^{\varrho\left(1-\frac{x}{r\sub{0}{}}\right)} - 2 \right)e^{\varrho\left(1-\frac{x}{r\sub{0}{}}\right)} & : x < \rcut \\
 0 & : \mathrm{otherwise}
\end{array} \right.
\label{eq:Morse}
\end{equation}

The capsomer-polymer interaction is a short-range repulsion that accounts for excluded-volume. For capsomer $i$ with excluder positions $\{\bm x_{i\alpha}\}$ and polymer subunit $j$ with position $\bm R_j$, the potential is:
\begin{eqnarray}
\label{Ucp}
U\sub{cp}{}(\{\bm x_{i\alpha}\}, \bm R_j) &=&
    \sum_{\alpha}^{N\sub{x}{}} \LJ{} \left(
    | \bm{x}_{i\alpha} - \bm R_j |,
    \sigma\sub{xp}{}\right)
\end{eqnarray}
where $N_\text{x}$ is the number of excluders on a capsomer and $\sigma\sub{xp}{} = 0.5(\sigma_\text{x}+\sigma_\text{p})$ is the effective diameter of the excluder -- polymer repulsion. The diameter of the polymer bead is $\sigma_\text{p}=0.5$ nm and the diameter for the excluder beads is $\sigma_\text{x}=3.0$ nm for the $T{=}1$ model and $\sigma_\text{x}=5.25$ nm for the $T{=}3$ model.

The capsomer-ARM interaction is a short-range repulsion that accounts for excluded-volume. For capsomer $i$ with excluder  positions $\{\bm x_{i\alpha}\}$ and ARM subunit $j$ with position $\bm R_j$, the potential is:
\begin{eqnarray}
\label{UcA}
U\sub{cA}{}(\{\bm x_{i\alpha}\}, \bm R_j) &=&
    \sum_{\alpha}^{N\sub{x}{}} \LJ{} \left(
    | \bm{x}_{i\alpha} - \bm R_j |,
    \sigma\sub{xA}{}\right)
\end{eqnarray}
with $\sigma\sub{xA}{} = 0.5(\sigma_\text{x}+\sigma_\text{A})$ as the effective diameter of the excluder - ARM repulsion with $\sigma_\text{A}=0.5$ nm the diameter of an ARM bead.

The polymer-polymer non-bonded interaction is composed of electrostatic repulsions and short-ranged excluded volume interactions. These polymers also contain bonded interactions which are only evaluated for segments occupying adjacent positions along the polymer chain and angular interactions which are only evaluated for three sequential polymer segments. As noted in the main text, electrostatics are represented either by Debye Huckel interactions or by Coulomb interactions with explicit salt ions. For the case of Debye Huckel interactions,

\begin{align}
\label{Upp}
U\sub{pp}{}(\bm R_i, \bm R_j, \bm R_k)
    &=&
    \left\{
            \begin{array}{l}
                \mathcal{K}_\mathrm{bond}(R_{ij}, \sigma\sub{p}{},k_\mathrm{bond})\\
                \quad:\{i,j\}\ \mathrm{bonded} \\
                \mathcal{K}_\mathrm{angle}(R_{ijk}, k_\mathrm{angle})\\
                \quad: \{i,j,k\}\ \mathrm{angle} \\
                \LJ{}(R_{ij}, \sigma\sub{p}{}) + \mathcal{U}\rsub{DH}(R_{ij},\qp,\qp, \sigma\sub{p}{})\\
                \quad: \{i,j\}\ \mathrm{nonbonded}\\
            \end{array}
        \right.
\end{align}
where $R_{ij} \equiv | \bm R_i - \bm R_j |$ is the center-to-center distance between the polymer subunits, $\qp=-1$ is the valence of charge on each polymer segment, and $\mathcal{U\rsub{DH}}$ is a Debye-Huckel potential smoothly shifted to zero at the cutoff:

\begin{align}
\mathcal{U}\rsub{DH}&(r,q_1,q_2, \sigma{}) \equiv \\
&\left\{  \begin{array}{l}
	\frac{q_1 q_2 l\rsub{b}\ e^{\sigma{}/\lambda\rsub{D}}}{\lambda\rsub{D}+\sigma{}} \left(\frac{e^{-r/\lambda\rsub{D}}}{r} \right)\\
	\quad: x < 2\lambda\rsub{D} \\
	 \frac{(r_{cut}^{2}-r^{2})^{2}(r_{cut}^{2}+2r^{2}-3r_{on}^{2}))}{(r_{cut}^{2}-2r_{on}^{2})^{3}} \frac{q_1 q_2 l\rsub{b}\ e^{\sigma{}/\lambda\rsub{D}}}{\lambda\rsub{D}+\sigma{}}
	\left(\frac{e^{-r/\lambda\rsub{D}}}{r} \right)\\
	\quad: 2\lambda\rsub{D} < x < 3\lambda\rsub{D} \\
	0 \\
	\quad: \mathrm{otherwise}
\end{array} \right.
\end{align}

$\lambda\rsub{D}$ is the Debye length, $l\rsub{b}$ is the Bjerrum length, and $q_1$ and $q_2$ are the valences of the interacting charges. For the cases using explicit electrostatics the $\mathcal{U\rsub{DH}}$ potential is replaced by a Coulomb potential:

\begin{equation}
\Coulomb{}(r,q_1,q_2) \equiv
\frac{q_1 q_2}{4 \pi \varepsilon_{0}\varepsilon_{r}}\frac{1}{r}
\label{eq:coulomb}
\end{equation}
where $4 \pi \varepsilon_{0}$ is the term for the permittivity of free space and $\varepsilon_{r}$ is the relative permittivity of the solution, set to 80.  Above a cutoff distance ($\rcut$ = 5 nm) the electrostatics are calculated using the particle-particle particle-mesh (PPPM) Ewald summation \cite{LeBard2012}. Explicit ions are included in these simulations to represent both neutralizing counterions and added salt. Ions interact with other charged beads in the solution according to the Coulomb potential (Eq.~\ref{eq:coulomb}) and interact with all beads through the repulsive shifted LJ interaction (Eq.~\ref{eq:LJ}).

Bonds are represented by a harmonic potential:
\begin{equation}
    \mathcal{K}_\mathrm{bond}(R_{ij}, \sigma, k_\mathrm{bond})
    \equiv
    	 \frac{k_\mathrm{bond}}{2}(R_{ij}-\sigma)^2
    \label{eq:bonds}.
\end{equation}
Angles are also represented by a harmonic potential:
\begin{equation}
    \mathcal{K}_\mathrm{angle}(R_{ijk}, k_\mathrm{angle})
    \equiv
    	 \frac{k_\mathrm{angle}}{2}(\vartheta_{ijk})^2
\end{equation}
where $\vartheta_{ijk}$ is the angle formed by the sequential polymer units $i,j,k$.

The ARM-ARM interaction is similar to the polymer-polymer interaction, consisting of non-bonded interactions composed of electrostatic repulsions and short-ranged excluded volume interactions. These ARMs also contain bonded interactions which are only evaluated for segments occupying adjacent positions along the polymer chain:

\begin{eqnarray}
\label{Uaa}
U\sub{aa}{}(\bm R_i, \bm R_j)
    &=&
    \left\{
            \begin{array}{l}
                \mathcal{K}_\mathrm{bond}(R_{ij}, \sigma\sub{a}{},k_\mathrm{bond})\\
                \quad: \{i,j\}\ \mathrm{bonded} \\
                \LJ{}(R_{ij}, \sigma\sub{a}{}) + \mathcal{U}\rsub{DH}(R_{ij},q_i,q_j, \sigma\sub{a}{})\\
                \quad: \{i,j\}\ \mathrm{nonbonded}\\
            \end{array}
        \right.
\end{eqnarray}
where $R_{ij} \equiv | \bm R_i - \bm R_j |$ is the center-to-center distance between the ARM subunits and $q_i$ is the valence of charge on ARM segment $i$.

Finally, the ARM-Polymer interaction is the sum of short-ranged excluded volume interactions and electrostatic interactions:

\begin{equation}
\label{Upa}
U\sub{pa}{}(\bm R_i, \bm R_j)
    =
	\LJ{}(R_{ij}, \sigma\sub{ap}{}) + \mathcal{U}\rsub{DH}(R_{ij},q_i,q_j, \sigma\sub{ap}{})
\end{equation}

\subsection{Base-paired polymer}
\label{sec:S1B}
\begin{figure*}
\centering{\includegraphics[width=0.99\textwidth]{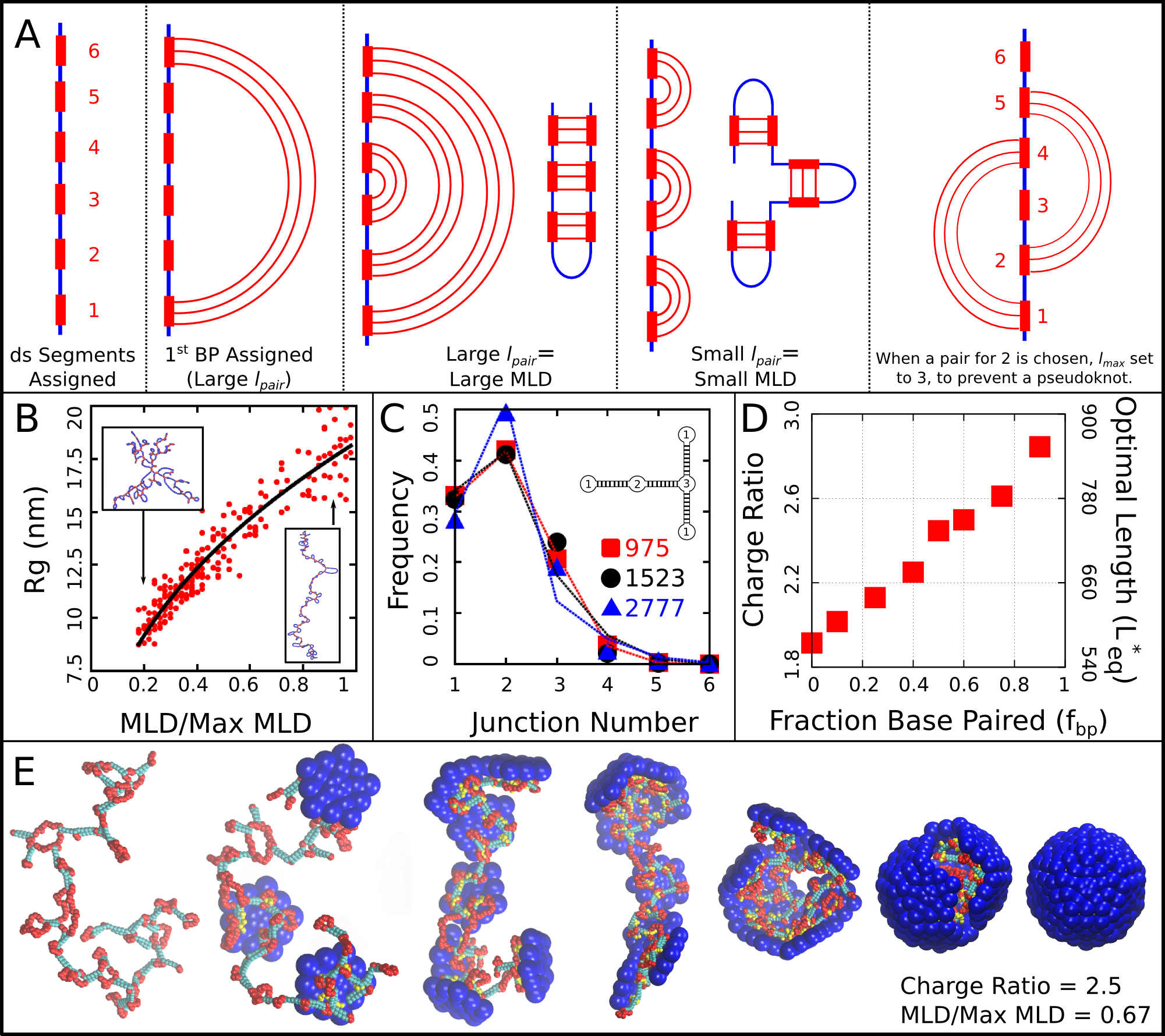}}
\caption{{\bf (A)} Schematic illustrations of the algorithm we used to obtain a wide range of base-paired structures. From left to right, double-stranded (ds) segments are first randomly assigned. These segments are then base-paired together, starting from one end. If base-paired segments are widely separated (i.e. $\lpair$ is large) then subsequent nested base-pairs lead to an extended structure. Conversely, if $\lpair$ is small, less extended structures form. The right-most panel indicates a psuedoknot, a structural motif we have prevented from occurring in this model, by setting $l_\text{max}$ to the last unpaired segment. {\bf (B)} Radius of gyration $\Rg$ for model NAs isolated in solution as a function of maximum ladder distance (MLD) normalized by the maximum possible MLD. The nucleic acid has 1000 nt, 50\% of which are base-paired. {\bf (C)} The frequency of junction numbers can be altered by varying $\lambda$ in Eq.~\ref{eq:lPair}, with large values of $\lambda$ leading to large values of $\lpair$. The symbols indicate the relative frequency of junction numbers for biological RNAs with indicated lengths, obtained from Ref.~\cite{Gopal2012}, and the lines are best fits to these distributions generated by varying $\lambda$. The inset illustrates several different junction orders. {\bf (D)} The thermodynamic optimum length measured for the simple model capsid as a function of the fraction of base-paired nucleotides $\fbp$ for a simplified `hairpins only' model (\textcolor{red}{$\blacksquare$} symbols). {\bf (E)} Snapshots illustrating assembly around a NA. Beads are colored as follows: blue=excluders, yellow=ARM bead, red=single-stranded NA, cyan=double-stranded NA. `Top', `Bottom', and `Attractor' beads removed for clarity.
\label{figS3}
}
\end{figure*}

To obtain base-paired polymers with a wide and tunable range of structures (i.e. maximum ladder distances), we implement the following strategy. Firstly, the polymer contour length $\Lc$, length of the base-paired segments $\Lbp$, and fraction of nucleotides in base-pairs $\fbp$ are free parameters which we specify (typical values are $\Lc=1000$ nucleotides, $\Lbp=5$ nucleotides per segment, and $\fbp=0.5$). Secondly, we iterate over the linear sequence of the polymer, randomly choosing segments which will undergo base-pairing to form double-stranded (ds) segments. Each segment consists of $\Lbp$ consecutive nucleotides. Segments are numbered sequentially to facilitate pairing (i.e. the first ds segment in the sequence is 1, the second is 2, and so on). Thirdly, these ds segments are then paired together. In the case of the hairpin model, each ds strand is paired with the next ds segment in the sequence (i.e. the first segment with the second, third with fourth, and so on). In the general base-pairing model, pairs are assigned stochastically according to an algorithm which allows us to simultaneously tune the distribution of junction orders and the maximum ladder distance (MLD). The algorithm is described in Figure~\ref{figS3}A and is defined as follows:

The first step in assigning a base-pair is to obtain a random separation $l_\text{random}$ from an exponential distribution where $\lambda$ is the inverse of the mean:
\begin{align}
\left(l_\text{random}(\lambda,l)\right) = \lambda e^{-l\lambda}
\label{eq:lRandom}.
\end{align}
To prevent pseudoknots this $l_\text{random}$ is then subtracted from the maximal available separation $l_\text{max}$ to yield $\lpair$:
\begin{align}
\lpair(l_\text{max}, l_\text{random}) = l_\text{max}-l_\text{random}
\label{eq:lPair}
\end{align}

The obtained $\lpair$ defines the number of segments separating the current segment from its base-pairing partner. With this algorithm, the single control parameter parameter $\lambda$ is used to control both the base-pairing pattern, and thus MLD and the distribution of junction types, i.e. the number of double stranded segments emerging from a single stranded intersection (see Fig.~\ref{figS3}C). When $\lambda$ is large, we are more likely to obtain small values of $l_\text{random}$, and thus large values of $\lpair$. Large $\lpair$ values lead to more extensive structures (i.e. larger MLDs and a larger fraction of 2-junctions). When $\lambda$ is lower, we have a broader distribution of $l_\text{random}$ values, and thus obtain smaller values of $\lpair$. If $\lpair$ is small, it creates higher-order junctions and regions which are not part of the MLD.

To describe the structures of the polymers generated by this algorithm, we make use of two structural parameters: the maximum ladder distance (MLD) and radius of gyration ($\Rg$). As in \cite{Yoffe2008}, we define the MLD as the largest number of base-pairs in any single path across the molecule's secondary structure. Figure~\ref{figS3}B describes the polymer radius of gyration $\Rg$ as a function of MLD, normalized by the maximal possible MLD (i.e. if all base-pairs were along a single path), for polymers of length 1000 with fraction base-pairing $\fbp=0.5$. All of the base-paired polymers are compressed relative to the linear polymer ($\Rg=25.5$ nm), but they differ amongst themselves significantly. We observe $\Rg$ to vary with MLD as $\Rg \sim \text{MLD}^{0.43}$ to yield sizes in the range $\Rg\approx8$ nm to $\Rg \approx20$.

\emph{Effect of MLD on optimal charge ratio.}
In order to estimate biological MLD values, we fit the histogram of junction numbers generated by our algorithm with different values of $\lambda$ and against the distribution of junction numbers obtained for biological ssRNA molecules in Ref. \cite{Gopal2012} (Fig.~\ref{figS3}C). For the two cellular, noncoding ssRNA segments, we obtain normalized MLDs of 0.55 and 0.36, and for a viral segment (RNA2 of CCMV) we obtain 0.25. As shown in Fig.~\ref{figS3}B the radii of gyration for RNAs with lengths of $\Lc=1000$ nt and the normalized MLDs of the cellular RNAs of 0.55 and 0.36 are respectively 14.1 nm and 11.8 nm. A 1000 nt RNA with the viral normalized MLD of 0.25 has $\Rg=10.1$ nm; i.e., the viral-like RNA is compressed by 14-29\% in solution. However, as shown in Fig.~\ref{fig3}C, the optimal charge ratios for these RNAs in the simple capsid model are within the large statistical error (we obtain 2.70, 2.75, and 2.78 respectively from a linear fit to the data).

\subsection{Modeling specific capsids}
\label{sec:specificCapsids}
\begin{figure}
\centering{\includegraphics[width=0.5\columnwidth]{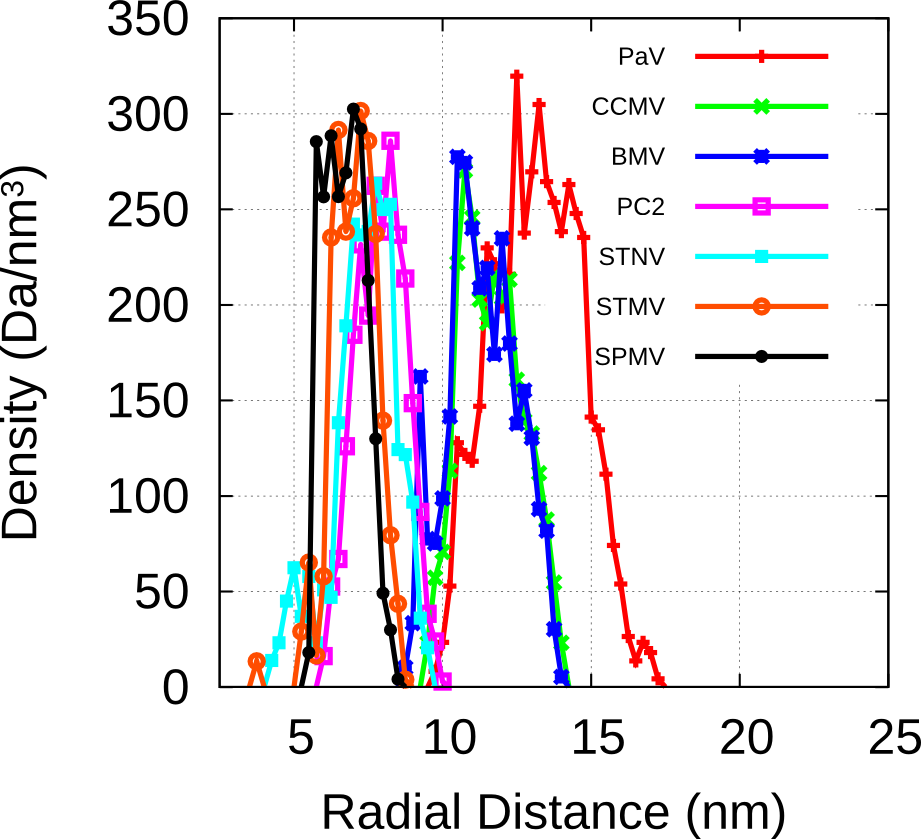}}
\caption{{\bf (A)} The radial density of capsid protein (C, N, S, O atoms) in biological capsid crystal structures, used to estimate capsid interior volumes to set the sizes of model capsids, as described in section~\ref{sec:specificCapsids}.}
\label{figS4}
\end{figure}

Our capsid model can be modified to describe specific viral capsids by altering the capsid radius and ARM sequence. Atomic-resolution structures of capsids are available for PC2, STNV, STMV, SPMV, PaV, BMV, and CCMV \cite{Khayat2011, Jones1984, Larson1998, Ban1995, Tang2001, Lucas2002, Speir1995}. For each capsid structure, we estimated the radius by fitting the radial density of capsid protein atoms (Figure~\ref{figS4}A) to a Gaussian. For $T{=}1$ capsids, we scaled the inradius of our dodecahedral model capsid (Fig.~\ref{fig1}) until its interior volume was equal to the volume of a sphere with the radius of the biological capsid. The ARMs were anchored as shown in Fig.~\ref{fig1}, midway across the pentagonal radius (we found that changing the locations of anchor points did not substantially affect $\Lopt$), and the sequence of positive, negative, and neutral beads was set to match the amino acid sequence of the capsid protein for the virus being modeled. For $T{=}3$ capsids, an icosahedrally symmetric capsid was designed with the excluders and ARMs placed based on the crystal structure of the Brome Mosaic Virus \cite{Lucas2002}. For other $T{=}3$ viruses the ARM sequence and capsid radius were adjusted. For the satellite viruses, there are basic residues located on the capsid inner surface (in addition to those found in the ARM); for each such residue a positive charge was rigidly fixed to the inner surface of the model capsid. No atomic-resolution structures for capsids of viruses in the Nanoviridae family are available, so the capsid radius for BBT was based on electron microscopy \cite{harding1991}.

\begin{figure}
\centering{\includegraphics[width=0.99\textwidth]{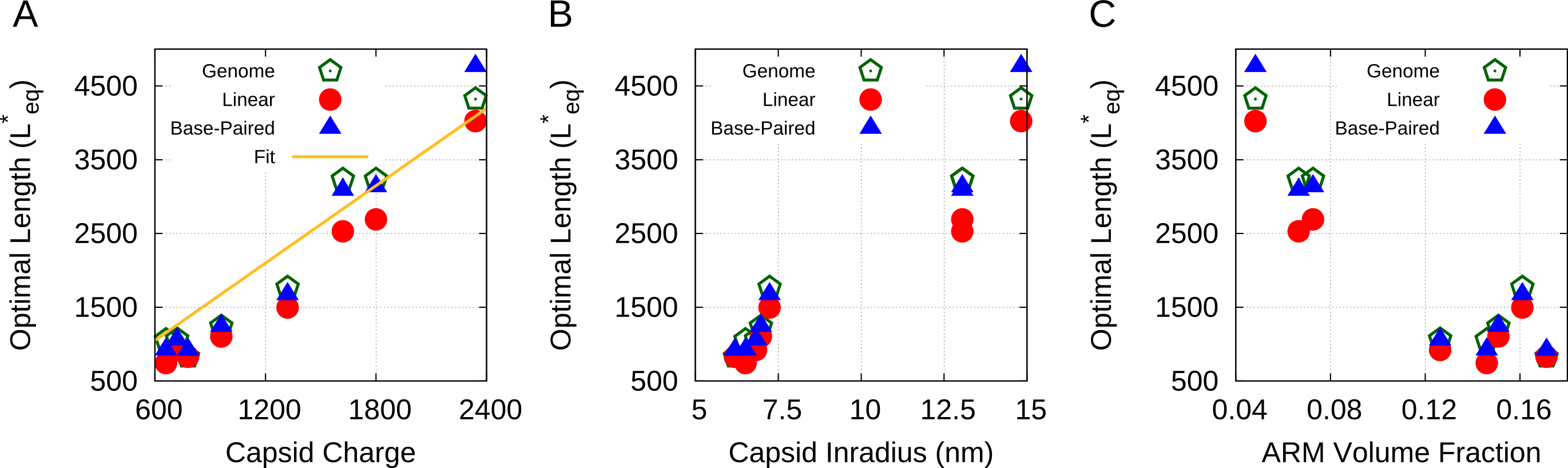}}
\caption{Values of the thermodynamic optimum length $\Lopt$ for the capsids considered in Figure~\ref{fig4} plotted against {\bf (A)} total capsid charge, {\bf (B)} capsid inradius, {\bf (C)} ARM volume fraction. Values are shown for a linear polyelectrolyte, the model base-paired NA, and the actual genome length for each virus.
\label{figS5}
}
\end{figure}

While Figure~\ref{fig4}B clearly shows that the optimal length cannot be uniquely related to the charge of the capsid, a correlation between capsid charge and genome length was previously noted \cite{Belyi2006}. A linear fit for the calculated values of $\Lopt$ for the viruses studied here yields a slope very similar to that previously observed (1.75), which we present in Figure~\ref{figS5}A as a comparison. We emphasize that the optimal length is also sensitive to other control parameters; e.g.,  $\Lopt$ is also correlated to capsid volume and ARM packing fraction (see Figures~\ref{figS5}B and \ref{figS5}C). Finally, we note that our simulations which include a base paired polymer provide a more accurate match to the biological genome lengths than those with a linear polyelectrolyte (Figure~\ref{fig4}), demonstrating the importance of the additional structural features considered in our model.

\section{Analysis of encapsidated polymer conformations}
\label{sec:bridging}

\begin{figure*}
\centering{\includegraphics[width=\textwidth]{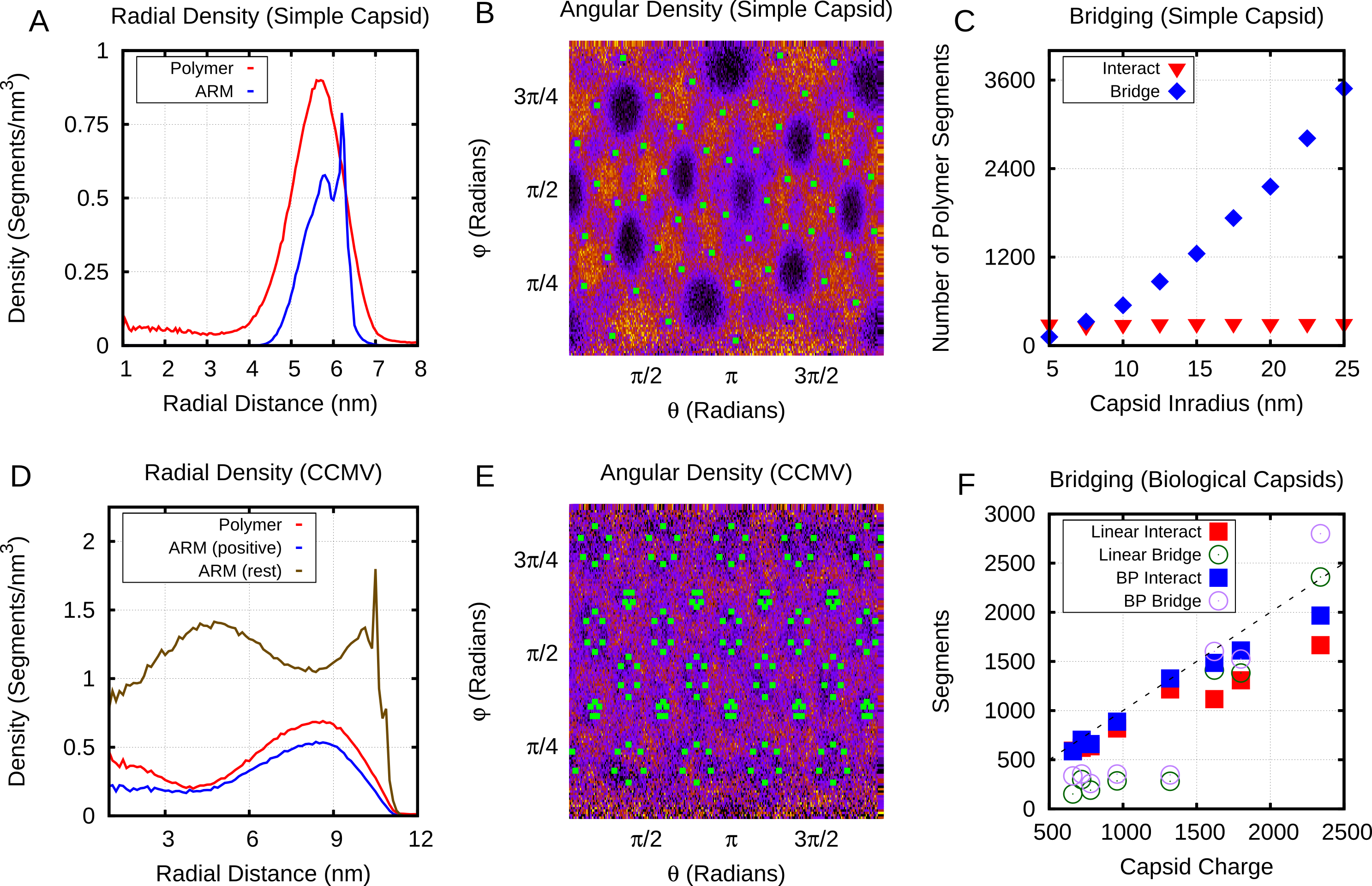}}
\caption{Structure of the encapsidated genome. {\bf (A,D)} Radial density for linear polymer and ARM segments in the simple capsid (A) and CCMV (D). {\bf (B,E)} Angular density of linear polymer segments (heat map) in the basic capsid model (B) and CCMV (E). Green squares indicate the first ARM segment. Segment densities are averaged over radial distances of 5-6.25 nm (B) and 8.75-10 nm (E), as a function of the spherical angles. {\bf (C)} Number of polymer segments interacting with positive capsid charges (\textcolor{red}{$\blacktriangledown$} symbols), and number of polymer segments not interacting with positive charges (bridging segments, \textcolor{blue}{\DiamondSolid} symbols). The numbers are calculated at the optimal polymer length $\Lopt$ as a function of capsid inradius $\inrad$ for the simple capsid with constant ARM length. {\bf (F)} Number of NA segments interacting with positively charged ARM segments and bridging. The numbers are calculated at the optimal length $\Lopt$ for each capsid using the linear and base-paired model. The 1:1 line is plotted for reference.}
\label{figS6}
\end{figure*}

Figure~\ref{figS6} describes the conformations of encapsidated polymers. Firstly, Fig.~\ref{figS6}A shows the radial density for linear polymer segments and ARM beads as a function of distance from the capsid center for our basic capsid model, where $\inrad=7.3$ nm. The sharp peak in ARM density is due to the first ARM segment, which is rigidly attached to the capsid shell. This figure shows that polymer segments are concentrated within a few nm of the capsid shell, with lower densities in the capsid center.

Figure~\ref{figS6}B presents the angular density distribution of linear polymer segments for the same capsid. The green squares indicate the position of the first ARM segments (which are rigid) and the heat map indicates the density of polymer over the course of the simulation (averaged over time and radial distances $5-6.25$ nm from the center in the capsid with $\inrad=7.3$nm, but with no angular averaging). This plot indicates that the polymer more frequently resides in the vertices between subunits (between the clusters of 3 ARMs) as well as along the dodecahedral edges, and resides less frequently in the center of the subunit faces.

In our particle-based model with discrete charges the internalized polymer must adopt distinct paths within the capsid, in which the polymer closely interacts with the ARM at some points, while in other places it merely bridges the gaps between interaction sites. To quantify bridging, we counted the number of polymer segments in close contact with ARM charges. We defined a threshold interaction distance of 0.74 nm, which encompasses the first peak in the positive charge - polymer segment radial distribution function and corresponds to a screened electrostatic interaction of $-0.5\kt$. We then found that for the basic capsid model ($\inrad=7.3$ nm and ARM length of 5 segments) less than half (44\%) of the polymer segments are strongly interacting (i.e. $<-0.5\kt$). The importance of the bridging segments is made particularly clear by the simulations in which capsid size was increased while maintaining fixed ARM length (Fig.~\ref{figS6}C). As shown in Fig.~\ref{figS6}C, the number of polymer segments strongly interacting with ARM charges is constant for $\inrad\ge 12.5$ nm, while the number of bridging segments increases to span the distances between arms. Hence, for capsids with $\inrad\ge12.5$ nm, the observed dependence of $\Lopt$ on capsid size arises entirely due to bridging segments. For smaller capsids, there is a weak increase in the number of interacting segments with size as more conformational space around the ARMs becomes available.

We would expect analysis of genomic configurations to be more complex for the simulations based on the specific biological capsid structures, since the longer arms result in a more diffuse distribution of positive charges within the capsid interior as compared to the basic capsid model. We begin by analysing the CCMV structure through a radial density distribution (Fig.~\ref{figS6}D), where we plot the density of the linear polymer, positive ARM segments, and combinined neutral and negative ARM segments. While there is some co-localization of positively charged ARM and negatively charged polymer segments, their densities peak at slightly different radii. The CCMV ARM sequence contains 48 segments, with 11 positive segments and 1 negative segment. Though the charges are not homogenously distributed throughout the sequence (9 occur within a 19 segment stretch), the degree of separation observed in Fig.~\ref{figS6}D was unexpected. Note that as in the Fig.~\ref{figS6}A, the sharp peak in the ARM distribution is due to the first immobilized ARM segment. As in the simple capsid model, the angular density is heterogeneous in CCMV, though to a lesser extent than found for the simple capsid(Fig.~\ref{figS6}E).

Bridging segments also make important contributions to overcharging in the biological capsids, and only when bridging segments are included are all of the biological capsids overcharged (see Fig.~\ref{figS6}F). Interestingly, we find that in $T{=}1$ capsids the fraction in contact is quite high (all above 60\%, with an average of 75\%) while in the $T{=}3$ capsids, the fraction in contact is substantially lower (all below 60\%, average of 44\%). It is not possible to directly quantify the fraction of bridging segments within crystallographic or EM structures since much of the RNA is not resolved and icosahedral symmetry is typically assumed. However, the predicted relationship between $T$ number and fraction of RNA segments which are strongly interacting with proteins is at least qualitatively suggested by the numbers of NA resolved in crystal structures. In a recent summary, Larsson et al. found that for 10 $T{=}3$ crystal structures an average of 16\% of NA were resolved. For $T{=}1$ structures a wider range of values was obtained, where some had a large fraction of NA resolved (STMV=62\%, STNV=34\%), but other ssDNA viruses had resolved fractions similar to $T{=}3$ viruses. An additional piece of evidence comes from low resolution neutron diffraction, where 72\% of RNA was observed to be in the first layer of density along the inner capsid surface of the $T{=}1$ STNV, again suggesting that much of the $T{=}1$ viral genome is closely interacting with the protein \cite{Bentley1987}. Finally, we note that Fig.~\ref{figS6}F shows that while the base-paired polymer increases the charge ratio it does so by increasing both the number of segments which are tightly bound and bridging.

\section{Additional calculations and analysis}

\subsection{Critical nucleus size}
\label{sec:kinetics}
\begin{figure}
\centering{\includegraphics[width=0.5\columnwidth]{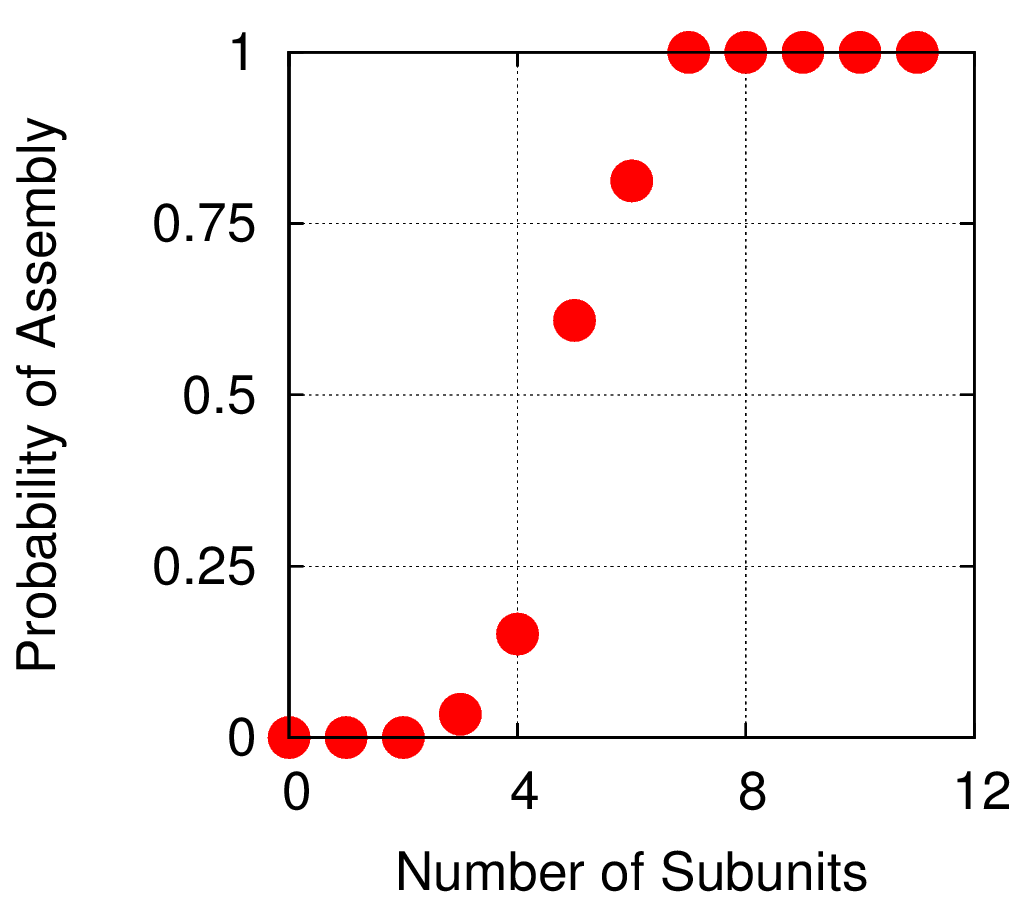}}
\caption{Commitment probability as a function of cluster size estimated from Brownian dynamics trajectories. The plot shows the measured probability of a cluster growing to completion before disassembling as a function of the largest cluster size, for trajectories with a linear polymer of length 575. The smallest cluster size with commitment probability exceeding 50\% is taken to be the critical nucleus size, $\nnuc=5$.}
\label{figS7}
\end{figure}

As in Ref. \cite{Kivenson2010}, we define the critical nucleus size $\nnuc$ as the smallest cluster of subunits from which more than 50\% of trajectories proceed to complete assembly before complete disassembly. To estimate $\nnuc$, we calculated cluster sizes for every recorded frame in our assembly trajectories, and recorded whether that cluster became part of a complete capsid before disassembling (defined as reaching a state of $n<3$). For the polymer length of 575, the critical nucleus size is $\nnuc=5$ (Fig.~\ref{figS7}).

\subsection{Subunit-subunit binding free energy estimates}
\label{sec:FreeEnergy}
Our method of calculating the subunit-subunit binding free energy is similar to that presented in our previous simulations \cite{Elrad2010, Hagan2011}. Briefly, for the simple capsid model (Fig. 1 of the main text), with an ARM length of 5, subunits were modified such that only one edge formed attractive bonds, limiting complex formation to dimers. We measured the relative concentration of dimers and monomers for a range of attraction strengths ($\varepsilon$). The free energy of binding along that interface is then $\gcc =-\kt \ln(c_\text{ss}/\kd)$ with standard state concentration $c_\text{ss}=1$ M and $\kd$ in molar units. This free energy is well fit by the linear expression $\gcc ~ -1.5\varepsilon - T\sbb$ where $\sbb=-9\kt$. We can then correct for the multiplicity of dimer conformations, by adding in the additional term $T-\Delta\scc = \ln(25/2)$, where the five pentagonal edges are assumed to be distinguishable, but complex orientations which differ only through global rotation are not. Our assembly simulations are run at $\varepsilon= 5\kt$, for which we observe only transient subunit-subunit associations except in the presence of a anionic polyelectrolyte. Our free energy calculations agree with this observation, suggesting that for this value of $\varepsilon$ binding is very weak: $\kd=0.33M$ and $\gcc=-1.1\kt$. Note that formation of additional bonds in a capsid structure will give rise to substantially more negative binding free energies. As shown in Ref.~\cite{Hagan2006} much of the binding entropy penalty associated with adding a subunit to a capsid is incurred during the formation of the first bond, with smaller decreases in entropy associated with forming additional bonds. A similar set of calculations for capsids with the ARMs removed decreased the binding free energy to $\gcc=-1.84 \kt$, indicating that ARM-ARM interactions increase the free energy by about $0.74\kt$ along each interface at $\Csalt=100$ mM.

\subsection{Equilibrium encapsidation}
\label{sec:equilibrium}
The free energy as a function of encapsidated polymer length was obtained by two different methods. In the first, we placed a very long polymer in or near a preassembled capsid, with one of the capsid subunits made permeable to the polymer. We then performed unbiased Brownian dynamics. Once the amount of packaged polymer reached equilibrium, the thermodynamic optimum length $\Lopt$ and the distribution of fluctuations around it were measured.

\begin{figure}[bt]
\centering{\includegraphics[width=0.5\columnwidth]{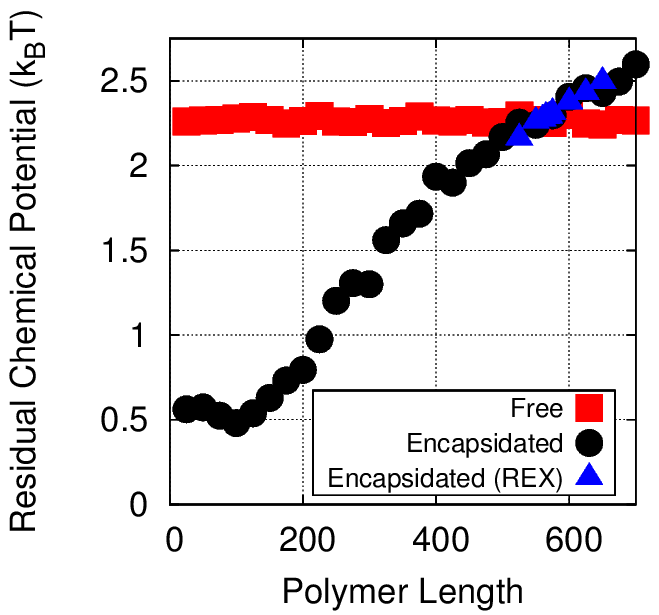}}
\caption{The residual chemical potential $\ur$ calculated by the Widom test-particle insertion method is shown for a linear polyelectrolyte, isolated in solution (\textcolor{red}{$\blacksquare$} symbols) and encapsidated in the simple capsid model (Fig.~\ref{fig1} main text) ($\bullet$ symbols). Results from replica exchange (REX) simulations on the encapsidated polymer are also shown (\textcolor{blue}{$\blacktriangle$} symbols).}
\label{figS8}
\end{figure}

In the second approach we used the Widom test-particle method \cite{Widom1963} as extended to calculate polymer residual chemical potentials \cite{Kumar1991, Elrad2010}. We performed independent sets of simulations for a free and an encapsidated polymer in which we calculated the residual chemical potential $\ur$ according to:

\begin{align}
-\beta\ur(\np) &\equiv
-\beta(\uchain(\np+1)-\uchain(\np)) \nonumber \\
&= \log<\exp(- \beta \ui (\np))>
\end{align}
where $\np$ is the number of segments in the polymer and $\ui$ is the interaction energy experienced by a test segment inserted onto either end of the polymer. Importance sampling was used to make the calculation feasible, where the bond length of inserted segments was chosen from a normal distribution matching the equilibrium distribution of bond lengths, truncated at $\pm 3$ standard deviations. The effect of using this biased insertion was removed \textit{a posteriori} according to standard non-Boltzmann sampling. Between incrementing $\np$, $10^{5}$ steps of dynamics were run, and $10^{5}$ insertions were attempted for each value of $\np$ in 100 independent runs. The results of these calculations are presented in Figure~\ref{figS8}, and based on the point of intersection between the encapsidated and unencapsidated chemical potentials, we estimate the optimal length $\Lopt$ to be between 550-575 segments (or a charge ratio of $1.83-1.92$), which is close agreement with the preassembled dynamics calculations (574 segments or a charge ratio of 1.91). If we integrate the difference in chemical potential between the encapsidated and unencapsidated polymers between $0$ and $575$, we obtain $-500 kT$ as an estimate for the free energy of polymer encapsidation due to both polymer-ARM and polymer-polymer interactions in the simple capsid model with ARM length=5. Since the primary motivation for the Widom test-particle method calculations was to provide an independent test of optimal lengths calculated using the semipermeable capsid, we only considered the Debye-Huckel model for electrostatics in test-particle method calculations.

To further assess the convergence and sampling of both approaches for calculating the $\Lopt$, we performed additional replica exchange (REX) simulations \cite{sugita1999}. In replica exchange, replicas of the system are simulated in parallel at different temperatures. Periodically, structures are exchanged between temperatures based on the Metropolis Criterion. In our systems, 12 replicas were run, with temperatures distributed exponentially between 1 kT and 1.5 kT. This resulted in a satisfactory exchange frequency of 30-40\%. We present the results for REX simulations in Figure~\ref{figS7}, but in this case and all other cases, the REX results quantitatively agree with the results of our simulations run at a single temperature.

\subsection{Linear charge density and condensed counterions}
\label{sec:CounterionCondensation}
In a previous work \cite{Belyi2006} it was noted that the high charge densities of RNA and peptide tails will give rise to counterion condensation, where the linear charge density of a polyelectrolyte is renormalized by condensed counterions to an effective linear charge density of $1/\lb$ charges/nm, with $\lb=0.714$ nm as the Bjerrum length. It was proposed that these renormalized charge densities should be used when calculating the electrostatic free energy for polyelectrolyte encapsidation. We chose not to do so in our simulations which used the Debye-Huckel model because association of RNA or an anionic linear polyelectrolyte with the oppositely charged peptide tails will lead to dissociation of condensed counterions. To test the validity of this choice (and to further test the validity of the Debye-Huckel model), we calculated optimal lengths $\Lopt$ as a function of the linear charge density for a linear polyelectrolyte using both the Debye-Huckel (with no assumed counterion condensation) and Coulomb interactions with explicit counterions. In the latter simulations counterion condensation arises naturally and responds to local charge densities with no approximations. The linear charge density was varied by adjusting the equilibrium bond lengths between neighboring beads in the polyelectrolyte; all other parameters were unchanged from those described in the main text.

\begin{figure}
\centering{\includegraphics[width=0.5\columnwidth]{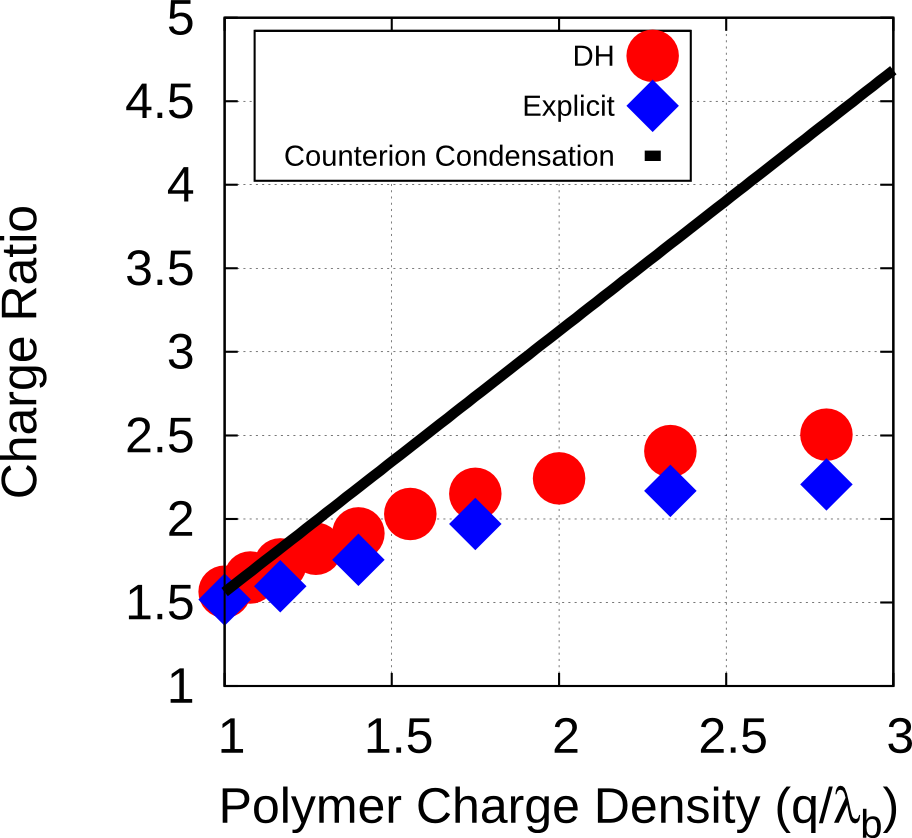}}
\caption{The optimal charge ratio as a function of polyelectrolyte linear charge density predicted by simulations in which electrostatics were calculated using Debye-Huckel (DH) electrostatics with no assumed counterion condensation (\textcolor{red}{$\bullet$} symbols) or Coulomb interactions with explicit counterions (\textcolor{blue}{\DiamondSolid} symbols). The simulations used the simple capsid model ($\inrad=7.3$, ARM Length = 5) at a Deybe length of $\ld=1$ nm or a counterion concentration of 100mM, respectively. The results are compared to the results of simulations with the Debye-Huckel model and irreversible counterion condensation (black line).}
\label{figS9}
\end{figure}

As shown in Fig.~\ref{figS9}, the optimal charge ratios calculated by the two methods qualitatively agree for all charge densities, although the quantitative agreement decreases as the charge density increases. The latter result was expected, since the accuracy of the Debye-Huckel approximation decreases with increasing local charge densities. Both models predict that the optimal charge ratio increases weakly with charge density, due to the fact that the excluded-volume per polyelectrolyte charge decreases. We also show the optimal charge ratios that would be predicted by the Debye-Huckel model with irreversibly renormalized charge (black line). To obtain this result, we performed a single simulation using the Debye-Huckel model with a linear charge density of 1 charge/$\lb$, and then assumed all charge densities exceeding this value are renormalized, so that the optimal charge ratio increases proportionally with the bare linear charge density. I.e. at a charge density of 2 charges/$\lb$, only half the polymer is effectively charged and the optimal charge ratio (calculated as a ratio of bare charge on the RNA to bare charge on the peptide arms) doubles. We see that the prediction of this scenario qualitatively disagrees with the explicit counterion results.

\end{document}